\documentclass[conference]{IEEEtran}
\IEEEoverridecommandlockouts
\usepackage{cite}
\usepackage{amsmath,amssymb,amsfonts}
\usepackage{algorithm}
\usepackage{algorithmic}
\usepackage{graphicx}
\usepackage{subcaption}
\usepackage{textcomp}
\usepackage{xcolor}
\usepackage{url}
\usepackage{booktabs}
\def\BibTeX{{\rm B\kern-.05em{\sc i\kern-.025em b}\kern-.08em
    T\kern-.1667em\lower.7ex\hbox{E}\kern-.125emX}}

\begin{document}

\title{A general-purpose hierarchical mesh partitioning method with node balancing strategies for large-scale numerical simulations}

\author{
  \IEEEauthorblockN{%
    Fande Kong\IEEEauthorrefmark{2}\IEEEauthorrefmark{1},
    Roy H.~Stogner\IEEEauthorrefmark{3},
    Derek R.~Gaston\IEEEauthorrefmark{2},
    John W.~Peterson\IEEEauthorrefmark{2},
    Cody J.~Permann\IEEEauthorrefmark{2},\\
    Andrew E.~Slaughter\IEEEauthorrefmark{2},
    and Richard C.~Martineau\IEEEauthorrefmark{2}}
  \IEEEauthorblockA{\IEEEauthorrefmark{2}Department of Modeling and Simulation\\
    Idaho National Laboratory, Idaho Falls, Idaho, USA}
  \IEEEauthorblockA{\IEEEauthorrefmark{3}Institute for Computational and Engineering Sciences\\
    The University of Texas, Austin, Texas, USA
  }
    \IEEEauthorblockA{\IEEEauthorrefmark{1}Corresponding Author: fande.kong@inl.gov}
}

\maketitle

\begin{abstract}
  Large-scale parallel numerical simulations are essential for a wide range of engineering problems
  that involve complex, coupled physical processes interacting across a broad range of spatial
  and temporal scales.
  The data structures involved in such simulations (meshes, sparse
  matrices, etc.) are frequently represented as graphs, and these graphs
  must be optimally partitioned across the available computational
  resources in order for the underlying calculations to scale
  efficiently.
  Partitions which minimize the number of graph edges that are cut
  (edge-cuts) while simultaneously maintaining a balance in the amount
  of work (i.e. graph nodes) assigned to each processor core are desirable,
  and the performance of most existing partitioning software begins
  to degrade in this metric for partitions with more than than $O(10^3)$ processor cores.
  In this work, we consider a general-purpose hierarchical partitioner which takes into account the existence of
  multiple processor cores and shared memory in a compute node
  while partitioning a graph into an arbitrary number of subgraphs.
  We demonstrate that our algorithms significantly improve the
  preconditioning efficiency and overall performance of realistic
  numerical simulations running on up to 32,768 processor cores with nearly $10^9$
  unknowns.
\end{abstract}


\section{Introduction}
As the compute node density, complexity, and heterogeneity of modern
supercomputers continues to advance, so too does the potential for
parallel numerical simulations to tackle ever more detailed and
demanding physical applications, and the need for sophisticated
numerical algorithms which are tailored to the underlying hardware.
The manner in which the data structures and workload of a given
simulation are partitioned among the available computational resources
plays a critical role in the overall parallel efficiency which can be
achieved, and the need to determine an ``optimal'' partitioning
becomes even more acute as the number of processor cores gets large.

In this work, we focus on developing a partitioning algorithm for
subdividing a large-scale graph (corresponding, e.g., to the dual graph
of a finite element mesh) into $np$ subgraphs of nearly equal size,
while minimizing the number of edge-cuts required to do so.
In this context, $np$ is the number of processor cores, the size of a
processor's subgraph is proportional to its computational workload
(thus, producing equal size subgraphs is sometimes referred to as
``load balancing''), and the number of edge-cuts corresponds to the
amount of inter-process communication required to share information
between parts of the graph. Computational workload calculations can
also be generalized to include the effects of so-called ``weighted''
graphs.  In a weighted graph, each vertex (and possibly each edge) is
assigned a ``weight'' value according to some metric, and the
corresponding workload is computed by summing the weights.  The
algorithm proposed in this work supports weighted graphs, but no
examples are considered here.

The computation of an optimal partitioning is known to be an
NP-complete problem \cite{andreev2006balanced}, so heuristics and
approximations must be employed to find acceptable partitions in a
reasonable time \cite{feldmann2015balanced}.
Interfaces between adjacent partitions result in shared degrees of
freedom (DOFs) which are communicated among neighboring partitions
during various parts of the simulation such as solver and preconditioner
applications.
Parallel scalability of solvers and preconditioners is a fundamental
requirement for constructing an efficient simulation capability, and
this scalability depends directly on partition quality.
Poor-quality partitions can lead to workload imbalances and
unnecessary communication, two major causes of parallel inefficiency.

Partitioning is an active research topic, and many approaches,
including geometric \cite{nour1987solving}, graph-based
\cite{farhat1988simple}, hierarchical~\cite{Devine_2002}, and spectral methods \cite{barnard1994fast}
have been proposed, studied, and implemented in numerical software
packages.  For larger processor counts, multilevel partitioning
frameworks \cite{karypis1998fast} have demonstrated reasonably good
results. The typical multilevel method consists of three stages:
coarsening, partitioning, and refinement. First, the original
large-scale graph is coarsened several times to reduce the graph
size. Next, the coarsest graph undergoes partitioning (typically using
a spectral method or graph-based algorithm). Finally, the partition is
improved during a refinement procedure, often using the local
optimization algorithms of Kernighan-Lin \cite{kernighan1970efficient}
(KL) and Fiduccia-Mattheyses \cite{fiduccia1982linear} (FM).

Based on the multilevel framework, several successful general-purpose
serial tools for graph partitioning have been developed, including
CHACO \cite{hendrickson1995chaco}, METIS \cite{karypis1995metis} and
Scotch \cite{chevalier2008pt}.
Most graphs of interest in large-scale numerical simulations are too
large to be partitioned on a single processor core, so distributed
memory parallel graph partitioning tools have also been developed,
including ParMETIS \cite{karypis1997parmetis} and PTScotch
\cite{chevalier2008pt}, which are based on underlying serial
multilevel partitioners.

Existing serial and parallel partitioners work well for up to
$O(10^3)$ processor cores, but they may be far from ideal or even fail to
generate partitions for larger numbers of processor cores.
Existing partitioners also frequently don't account for the layout
of the processing cores on multi-core systems by assigning neighboring
partitions to cores on the same compute node, an approach that can benefit
from memory locality and avoiding off-node communications entirely.

To address this shortcoming, we develop a hierarchical partitioning
approach which accounts for the existence of multiple cores per
compute node which is standard on modern supercomputers.
A similar idea has also been discussed in our previous work
\cite{Kong_2016, kong2016highly, kong2017scalable, kong2018efficient,
kong2016parallel}.

In the present study, the hierarchical partitioning approach is
extended to more general cases involving an arbitrary number of
submeshes, while still taking into account the number of cores
per compute node and the shared memory configuration.
If the partitioner is run as a preprocessing step,
the proposed algorithm can use a small number of processor cores to
partition a graph into a large number of subgraphs.
In addition, our new algorithm
ensures that the shared mesh nodes are evenly distributed among the
neighboring processor cores, thereby improving load balancing and
ultimately the overall parallel efficiency of the simulation.  The new
algorithm is referred to as the ``general-purpose hierarchical mesh
partitioning approach'' and abbreviated herein as ``Hierarch''.

The rest of this paper is organized as follows. In Section 2, the
general-purpose hierarchical mesh partitioning approach is described
in detail.  The various node assignment strategies are described in
Section 3, and numerical results with up to 32,768 processor cores  are
given in Section 4. The results are summarized and conclusions are drawn in Section 5.

\section{Hierarchical mesh partitioning}
\begin{figure}
 \centering
 \includegraphics[width=0.49\linewidth]{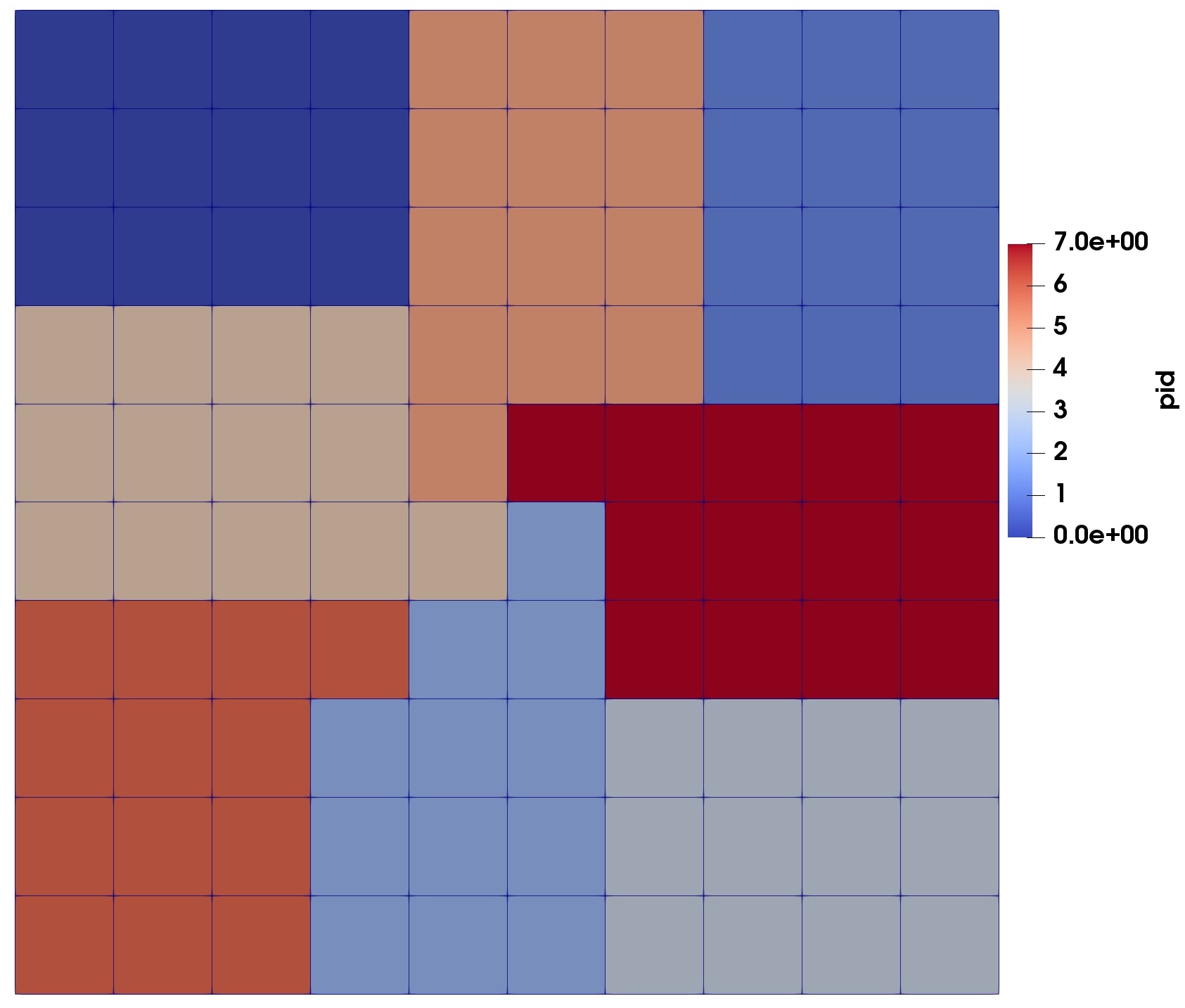}
 \includegraphics[width=0.49\linewidth]{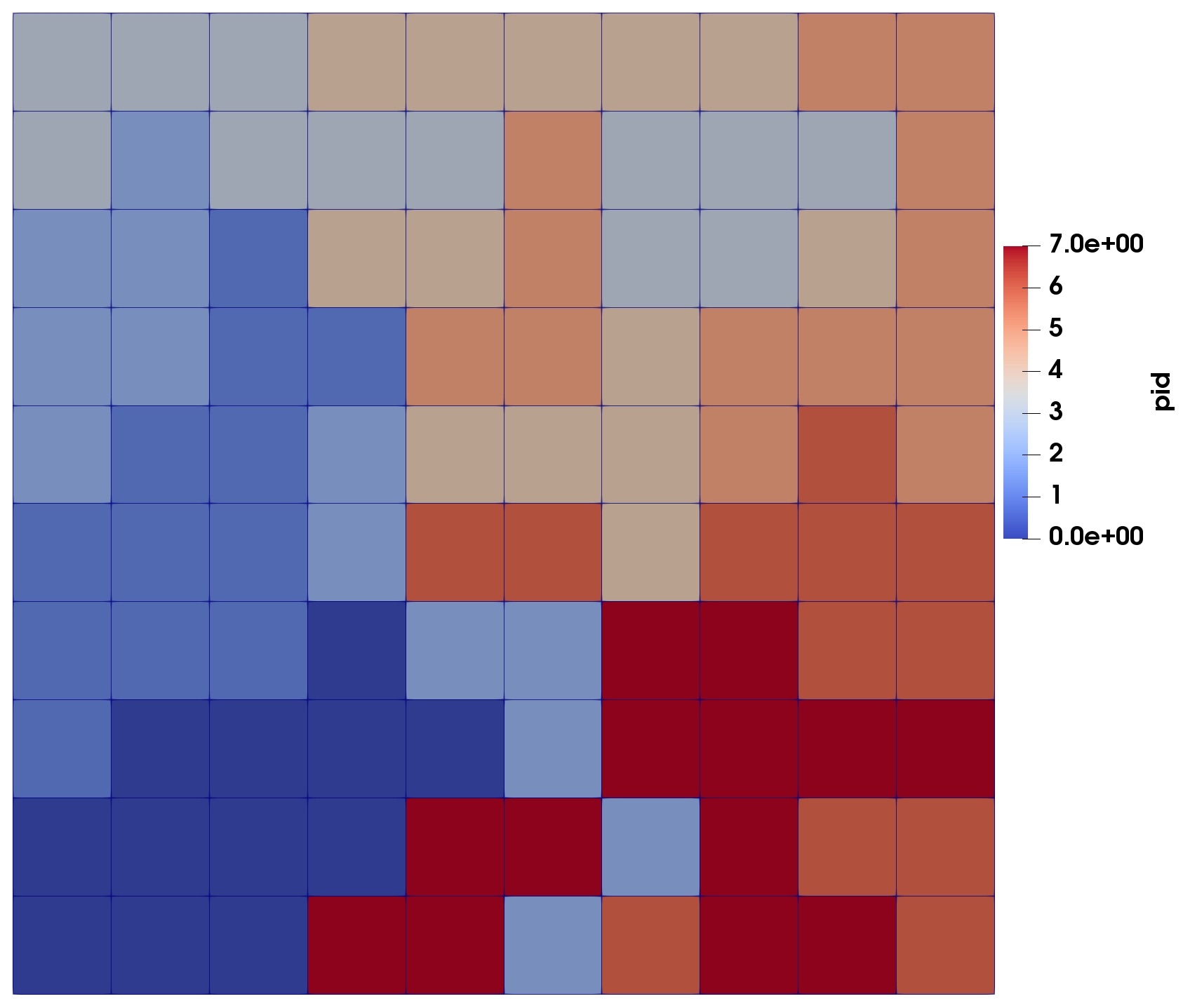} 
 \caption{``Good'' (left) and ``bad'' (right) examples of partitioning onto 8 subdomains. Each color corresponds to a different subdomain.\label{fig:good_bad_partitition_examples}}
\end{figure}

To further motivate the need for optimal partitioning and the
consequences of a poor partition, we first consider the simple 2D
example shown in Fig.~\ref{fig:good_bad_partitition_examples}.
In this example, a mesh is distributed across 8 different processors
using two different partitioning algorithms, and, upon inspection of
the results, it is clear (even with the small size of the problem) that
the result of the first partitioner is somewhat reasonable while
the quality of the second partition is quite poor.
To quantify the partition quality, the number of elements, nodes, and
edge-cuts for each processor is summarized in
Table~\ref{tab:good_bad_examples}. The number of the edge-cuts for the
``bad'' partition is clearly much larger than that for the ``good''
partition, and the workload, as measured by the number of nodes and
elements in each partition, is much more evenly split for the ``good''
partition than it is for the ``bad'' one.

\begin{table}
\scriptsize
\centering
\caption{2D partitioning example.  ``pid'' is the MPI rank, ``elems'' is the number of elements, ``nodes'' is the number of nodes, and ``edge-cuts'' is the number of edge-cuts.\label{tab:good_bad_examples}}
\begin{tabular}{c c c c c c  c c c c}
  \toprule
\multicolumn{4}{c}{Good partition example} \\
  \midrule
pid  &elems  & nodes & edge-cuts  \\
  \midrule
0 & 12 & 20 & 7 \\
1 & 12 & 20 & 7 \\
2 & 12 & 21 & 13 \\
3 & 12 & 16 & 7 \\
4 & 13 & 14 & 13 \\
5 & 13 & 10 & 13 \\
6 & 13 & 12 & 9 \\
7 & 13 & 8  & 13 \\
\bottomrule
\end{tabular}
~
\begin{tabular}{c c c c c c  c c c c}
  \toprule
\multicolumn{4}{c}{Bad partition example} \\
  \midrule
pid  &elems  & nodes & edge-cuts  \\
  \midrule
0 & 12 & 22 & 13 \\
1 & 12 & 17 & 17 \\
2 & 13 & 18 & 30 \\
3 & 12 & 21 & 21 \\
4 & 13 & 14 & 29 \\
5 & 13 & 11 & 23 \\
6 & 12 & 15 & 24 \\
7 & 13 & 3  &  23 \\
\bottomrule
\end{tabular}
\end{table}

The basic idea of hierarchical graph partitioning, which has been around
for some time (see, e.g., the Zoltan~\cite{Devine_2002} library in Trilinos), is to recursively
partition a graph two or more times, possibly onto different numbers
of processors each time.  For example, the graph may be initially
partitioned into $np_1$ subgraphs (where $np_1$ is, say, the number of
compute nodes) using an existing partitioner such as ParMETIS,
PTScotch, etc. Each subgraph can then be further partitioned into
$np_2$ smaller subgraphs (where $np_2$ is e.g.\ the number of
processor cores per compute node), so that the final partition has $np
= np_1 \times np_2$ partitions.

\begin{figure}
 \centering
 \includegraphics[width=0.49\linewidth]{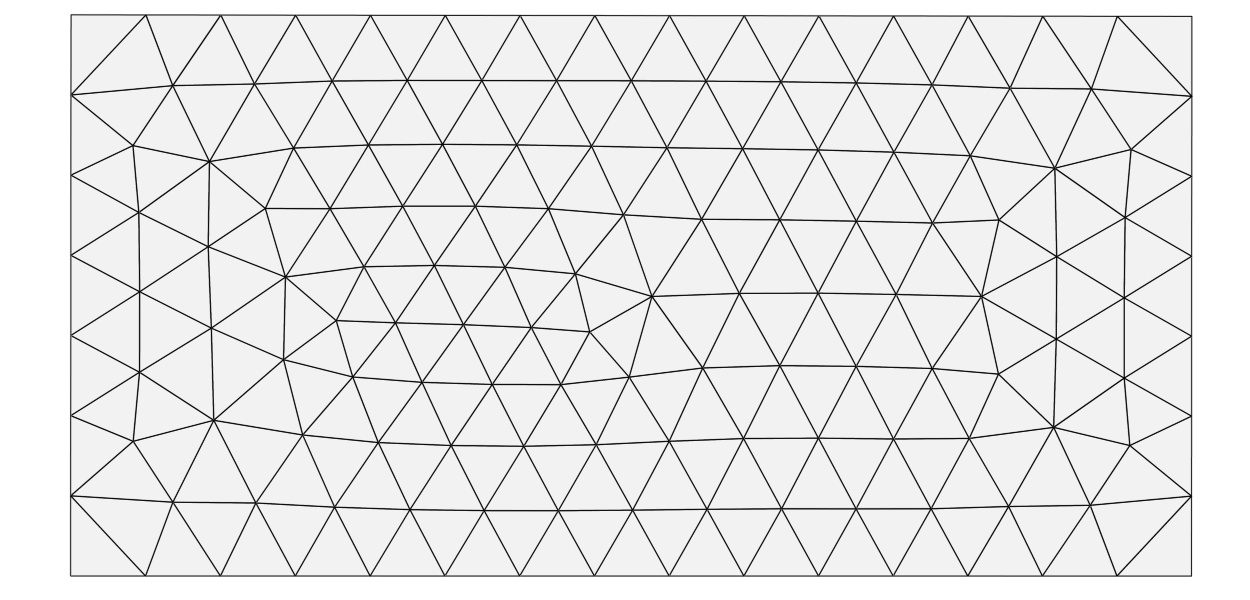}
 \includegraphics[width=0.49\linewidth]{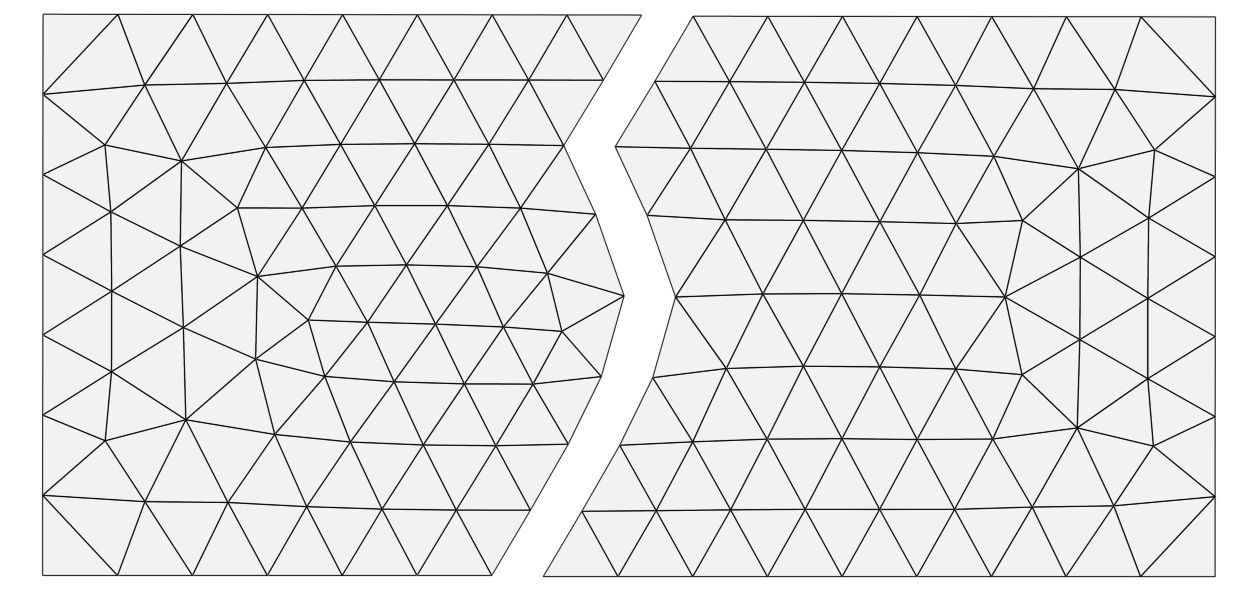} \\
  \includegraphics[width=0.49\linewidth]{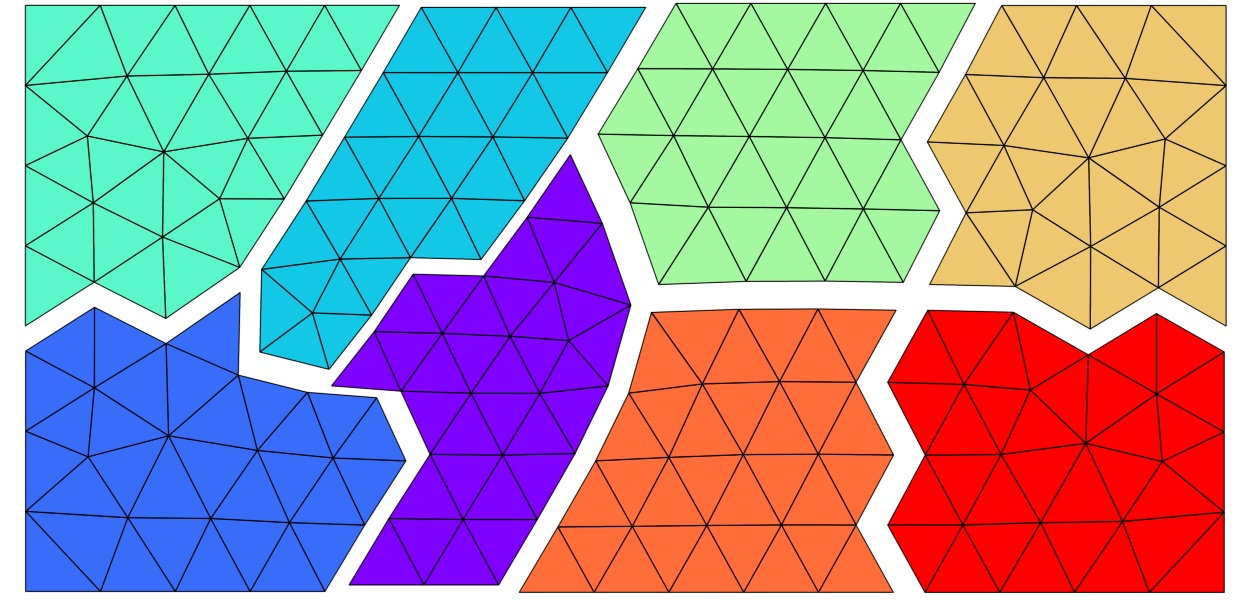}
  \caption{A 2D hierarchical partitioning example.  The original mesh
    in the top left is partitioned into the two submeshes shown in the top
    right, and each sub-mesh is further partitioned into the 4 small
    submeshes shown in the second row.\label{fig:hierarch_2D}}
\end{figure}

\begin{figure}
 \centering
 \includegraphics[width=0.49\linewidth]{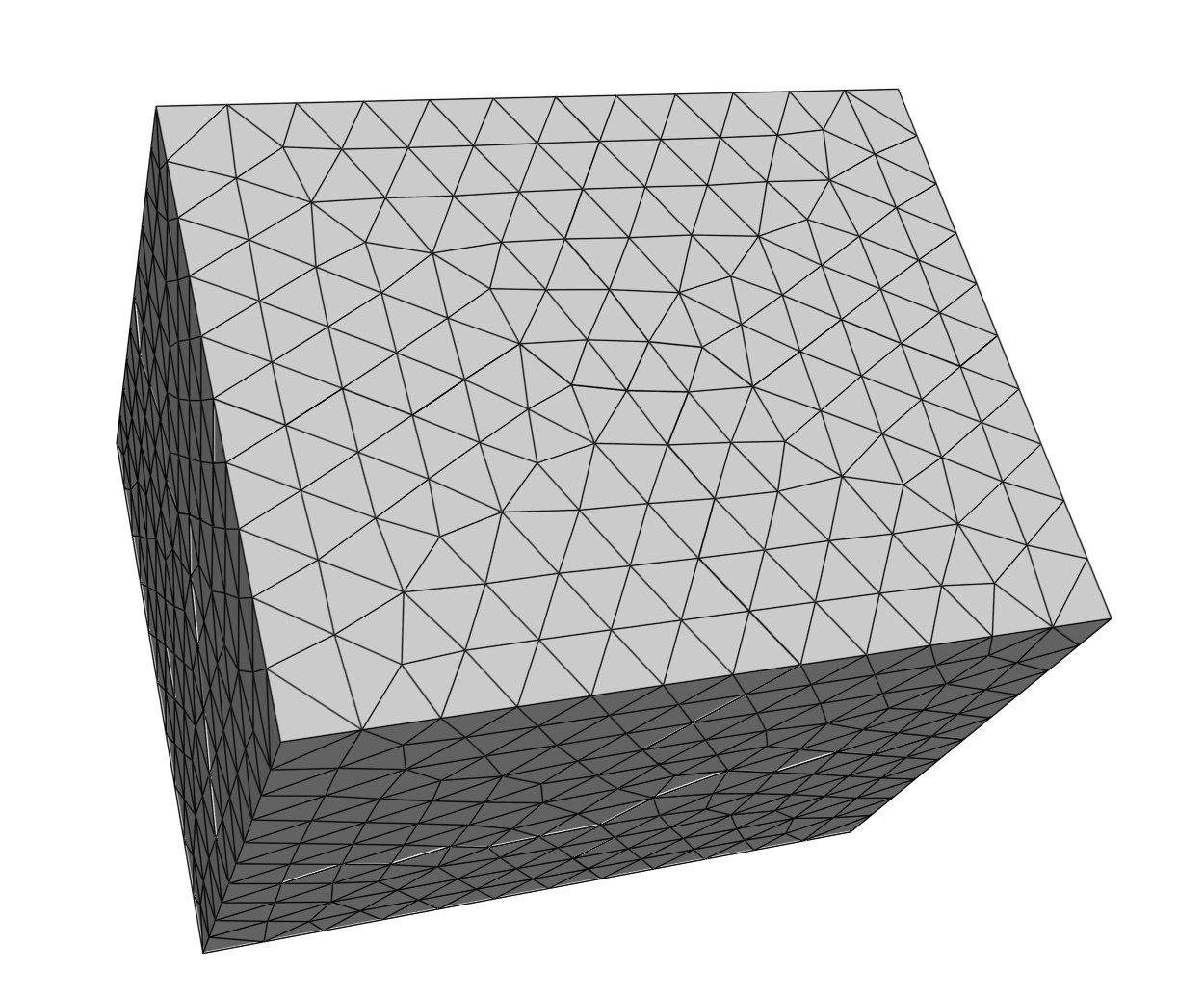}
 \includegraphics[width=0.49\linewidth]{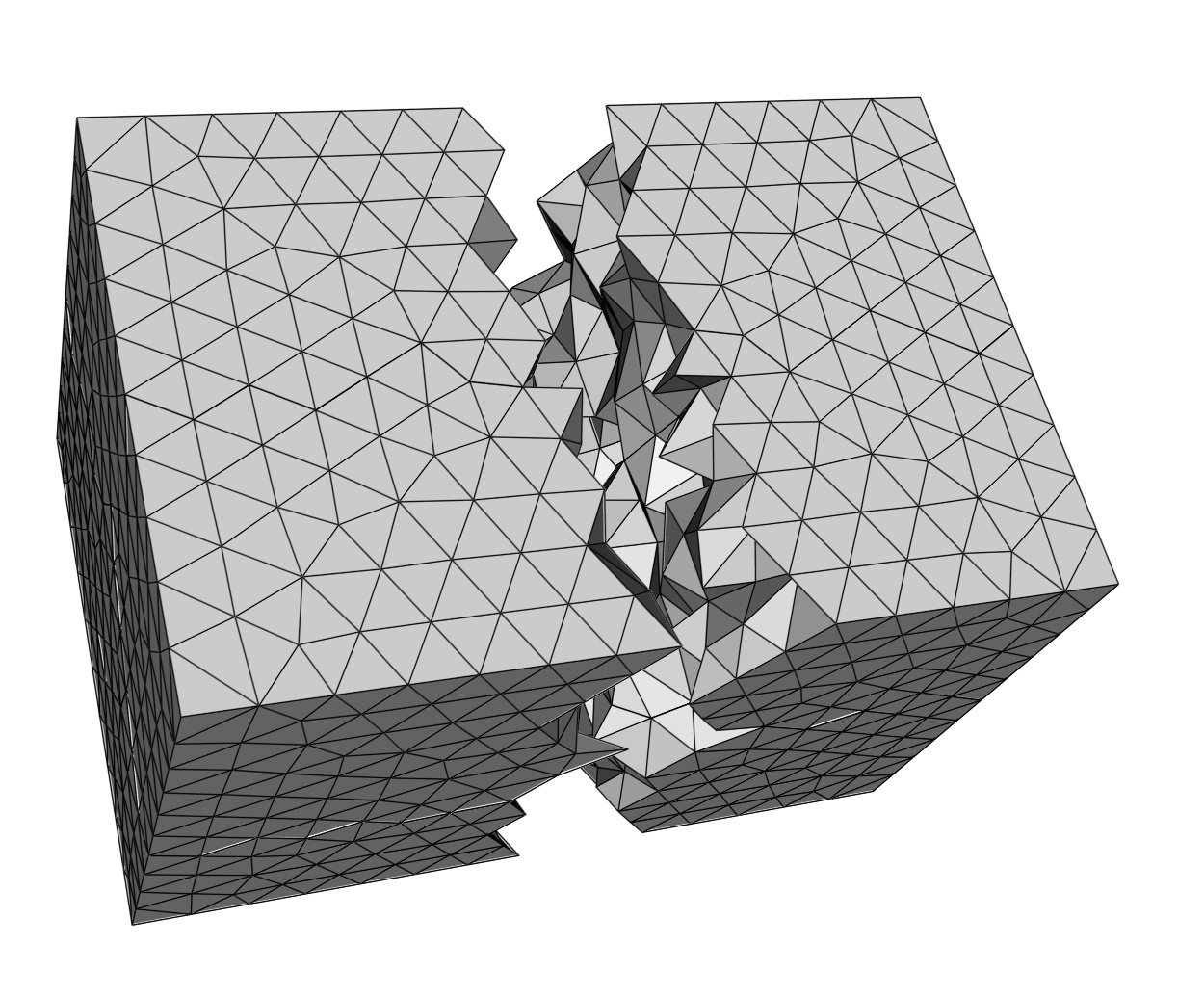} \\
  \includegraphics[width=0.49\linewidth]{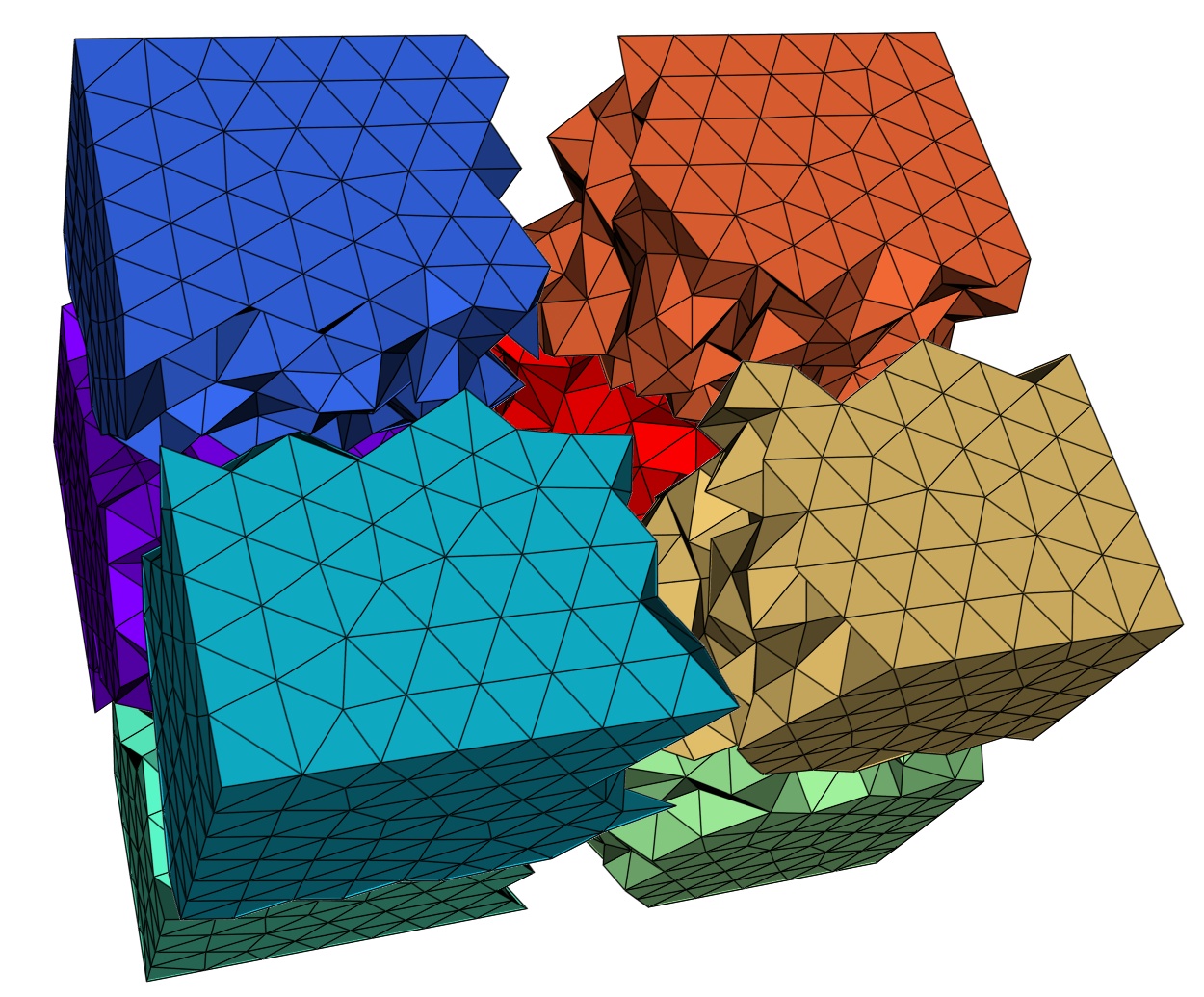}
  \caption{A 3D hierarchical partitioning 3D example.  As in the 2D case, the
    3D mesh is partitioned into 2 submeshes at the first step, and then
    each submesh is subsequently split into 4 smaller submeshes.\label{fig:hierarch_3D}}
\end{figure}

The basic idea is simple but very effective, especially when the number of
processor cores is large relative to the number of compute nodes, as
the likelihood of the algorithm failing increases with the number of
partitions asked for in a single iteration. If the number of compute nodes 
is itself large, the first step of ``Hierarch'' can be obtained in a recursive
manner. For simplicity, only the ``two-level'' hierarchical
partitioning method is discussed in this work. Two hierarchical
partitioning examples are shown in Figs.~\ref{fig:hierarch_2D}
and~\ref{fig:hierarch_3D}. In these cases, the original meshes are
initially partitioned into 2 submeshes, and then each submesh is
further split into 4 smaller submeshes. In total, 8 submeshes are
obtained through the hierarchical partitioning algorithm, and they can
be assigned to processor cores in such a way that adjacent partitions
are computed by processor cores on the same compute node.

To fix ideas and assist in describing the algorithm, let us 
denote the dual graph of the mesh as $G=(\{v_i\}, \{e_j\})$, where
$v_i$ is a graph vertex corresponding to a mesh element, and $e_j$ is
a graph edge representing a mesh element side. The number of graph
vertices $\{v_i\}$ is denoted by $nv$, and a partition is represented
by an integer array $P=\{ p_i \}$, $p_i \in [0, np)$, of size
$nv$. Vertices $v_i$ and $v_j$ are assigned to the same partition if and
only if $p_i = p_j$. The number of times $p_i$ is repeated indicates
how many vertices are assigned to partition $p_i$. To carry out the
partitioning scheme in parallel, the graph $G$ is assumed to be
initially distributed across the processor cores. This can be
accomplished by computing e.g.  the ``trivial'' partitioning defined
by assigning the first chunk of vertices (ordered by vertex ID) to the
first processor, the second chunk to the second processor, and so
on.

In the first step of the hierarchical partitioning scheme, the
distributed graph $G$ is partitioned by applying an existing
algorithm such as ParMETIS or PTScotch to produce a partition $P^1 =
\{ p^1_i\}$, $p^1_i \in [0, np_1), i=0, 1,\ldots, nv-1$. In order to
carry out the second partitioning step, $np_1$ subgraphs have to be
assembled and allocated to the first $np_1$ processors, with
each processor taking one subgraph. If the partitioning process is
carried out as a preprocessing step, the number of processor
cores used in the second step can be smaller than $np_1$. If this happens,
more than one subdomain will be allocated to the same processor core.
We do not discuss this situation in the present work since it is straightforward
to extend the algorithm to handle it, but our
implementation in \texttt{PETSc} does support this use case.

Two steps are required to construct a local graph from $P^1$. The
information in $P^1$ tells us where we should send the vertex IDs,
i.e. $v_i$ should be sent to the $p^1_i$th processor, but the required
data can't be exchanged in a single communication step. Instead,
communication ranks and data sizes have to be discovered before the
vertex IDs can be sent. The two-sided information discovery
operation is carried out using the algorithm discussed in
\cite{hoefler2010scalable}.

After the discovery, each processor knows how much information it will
receive and from whom it will receive it. The one-to-many sparse
communication pattern is efficiently implemented by the ``star
forest'' communication object in PETSc called \texttt{PetscSF}
\cite{brown2011star}. This vertex ID exchange algorithm is summarized
and implemented in the \texttt{ISBuildTwoSided} routine in PETSc.

Once the vertex exchange is complete, a vertex set $V_{c}$ is created
for the $c$th processor, where $c \in [0, np_1-1]$. A subgraph
$G_c$ is extracted from the global graph $G$ through, once again, a sparse
communication. The size of $G_c$ is denoted as $nv_c$. Finally, a serial partitioner (or parallel partitioner
on a single processor) is applied to
partition $G_c$ to produce $\tilde{P}^2_c = \{ p_{c,k}^2\}, p_{c,k}^2 \in [0, np_2-1), k=0, 1,\ldots, nv_c-1, c =0,1,\ldots, np_1-1$. The $\tilde{P}^2_c$ are sent
back to the original owners, and then are merged based on the global
vertex IDs. The new second-step partition is denoted as $P^2 = \{p^2_i\}, p^2_i \in [0, np_2-1), i = 0, 1,\ldots, nv-1$, the
entries of which are a permutation of $\{p_{0,0}^2, p_{0,1}^2,\ldots, p_{0,nv_0-1}^2,\ldots,   p_{c,0}^2, p_{c,1}^2,\ldots, p_{c,nv_{c}-1}^2,\ldots\}$. The
final partition $P$ is defined as follows:
\begin{equation*}
  P = \{ p_i = p^1_i \times np_2 + p_i^2 \},\, p^1_i \in P^1,\, p^2_i \in P^2.
\end{equation*}
The entire process is summarized in Algorithm~\ref{alg:hierarch},
where we assume $G$ is initially distributed across the processors.

\begin{algorithm}
\caption{Parallel hierarchical graph partitioning.\label{alg:hierarch}}
\begin{algorithmic}[1]
\STATE{Input: $G$}
\STATE{Partition $G$ into $np_1$ subgraphs denoted by
  $P^1 = \{ p^1_i \}$, $p^1_i \in [0, np_1),\, i=0, 1, \ldots, nv-1$,
  using a parallel partitioner such as ParMetis or PTScotch}
\STATE{Construct two-sided information for exchanging vertex IDs using a sparse discovery algorithm}
\STATE{Exchange vertex IDs and gather a new vertex ID set $V_c,\, c=0, 1,\ldots, np_1-1$}
\STATE{Extract a local subgraph $G_c$ from $G$ based on $V_c$}
\STATE{Apply a serial partitioner or parallel partitioner to $G_c$ to produce $\tilde{P}^2_c = \{ p_{c,k}^2\}, p_{c,k}^2 \in [0, np_2-1), k=0, 1,\ldots, nv_c-1, c =0,1,\ldots, np_1-1$}
\STATE{Send the entities of $\tilde{P}^2_c$ to their corresponding owners}
\STATE{Form the second-step partition, $P^2$, as a permutation of $\{p_{0,0}^2, p_{0,1}^2,\ldots, p_{0,nv_0-1}^2,\ldots,   p_{c,0}^2, p_{c,1}^2,\ldots, p_{c,nv_{c}-1}^2,\ldots\}$}
\STATE{Construct final partition
  $P = \{ p_i = p^1_i \times np_2 + p_i^2 \},\,
  p^1_i \in P^1,\,
  p^2_i \in P^2 $}
\STATE{Output: $P$}
\end{algorithmic}
\end{algorithm}

For some applications we need an arbitrary number, $np$, of
subgraphs, but $np$ cannot be factored into $np_1 \times np_2$.
Algorithm~\ref{alg:hierarch} has to be adjusted to handle this
situation, and our basic approach is to apply subdomain weights in the
first step so that a subgraph may have a different size from other
subgraphs, depending on how many smaller subgraphs it will be
further divided into. The remainder $r = np \mod np_2$ is computed
for a user-specified $np_2$ (usually $np_2$ is the number of processor
cores per compute node), and $np_1$ is calculated according to:
\begin{equation*}
  np_1 =
  \left \{
    \begin{array}{rl}
      np / np_2, &\text{if~} r = 0 \\
      np/np_2 +1, &\text{otherwise}.
    \end{array}
  \right .
\end{equation*}

The number of small subgraphs for each subgraph, $S=\{s_c\},
c=0,1,\ldots, np_1-1$, is constructed as follows:
\begin{equation*}
  s_c =
  \left \{
    \begin{array}{rl}
      r , &\text{if~} c=0 \text{~and~} r \neq 0 \\
      np_2, &\text{otherwise}.
    \end{array}
  \right .
\end{equation*}
Subdomain weights are simply derived from $S$ as $W={w_c = s_c/np}, c
=0, 1,\ldots, np_1-1$. A smaller weight indicates that less vertices
are assigned to the corresponding subgraph. An offset array, $O=\{ o_c
\}, c=0, 1,\dots, np_1-1$, of size $np_1$ is needed for forming the
final partition, and it is constructed by accumulating the entities of
$S$, that is,
\begin{equation*}
  o_c =
  \left \{
    \begin{array}{rl}
      0,           &\text{if~} c = 0, \\
      s_c + o_{c-1}, &\text{otherwise}.
    \end{array}
  \right.
\end{equation*}
In the second phase of the hierarchical partitioning, the subgraph $G_c$ is partitioned into $s_c$ small subgraphs, and the corresponding partition is denoted as $\tilde{P}^2_c$. Note that it is different from the previous algorithm where the $G_c$ is always partitioned into $np_2$ smaller subgraphs. The second step partition is formed similarly as ${P}^2$.   The final partition, $P=\{p_i\}$, is formed as
\begin{equation*}
p_i =  o_{p^1_i}+ p_i^2,  ~p^1_i \in P^1,\, p^2_i \in P^2. 
\end{equation*}
The generalized parallel hierarchical graph partitioning procedure is
summarized in Algorithm~\ref{alg:hierarch2}. The algorithm was
implemented in \texttt{PETSc} \cite{petsc-user-ref} as part of this work, and can be
accessed through the command-line option: \texttt{-mat\_partitioning\_type
hierarch}.
\begin{algorithm}
\caption{General-purpose  parallel hierarchical graph partitioning.\label{alg:hierarch2}}
\begin{algorithmic}[1]
\STATE{Input: $G$}
\STATE{Compute $W=\{w_c\}$, $O=\{o_c\}$ and $S=\{s_c\}$ }
\STATE{Partition $G$ into $np_1$ subgraphs denoted by
  $P^1 = \{ p^1_i \}$, $p^1_i \in [0, np_1), i=0, 1,\ldots, nv-1$,
  using a parallel partitioner such as ParMetis or PTScotch together with $W$}
\STATE{Construct two-sided information for exchanging vertex IDs using a sparse discovery algorithm}
\STATE{Exchange vertex IDs and gather a new vertex ID set $V_c, c=0, 1,\ldots, np_1 -1$}
\STATE{Extract a local subgraph $G_c$ from $G$ based on $V_c$}
\STATE{Partition $G_c$ to $s_c$ small subgraphs using a serial partitioner or parallel partitioner and then produce $\tilde{P}^2_c$}
\STATE{Send the entities of $\tilde{P}^2_c$ to their corresponding owners}
\STATE{Form second-step partition, $P^2$, as a permutation of $\bigcup_c^{np_1-1}\tilde{P}^2_c$}
\STATE{Construct final partition
  $P = \{ p_i = o_{p^1_i} + p_i^2  \},\,
  p^1_i \in P^1,\,
  p^2_i \in P^2,\, i=0, 1,\ldots, nv-1 $}
\STATE{Output: $P$}
\end{algorithmic}
\end{algorithm}

\begin{figure}
 \centering
 \includegraphics[width=0.49\linewidth]{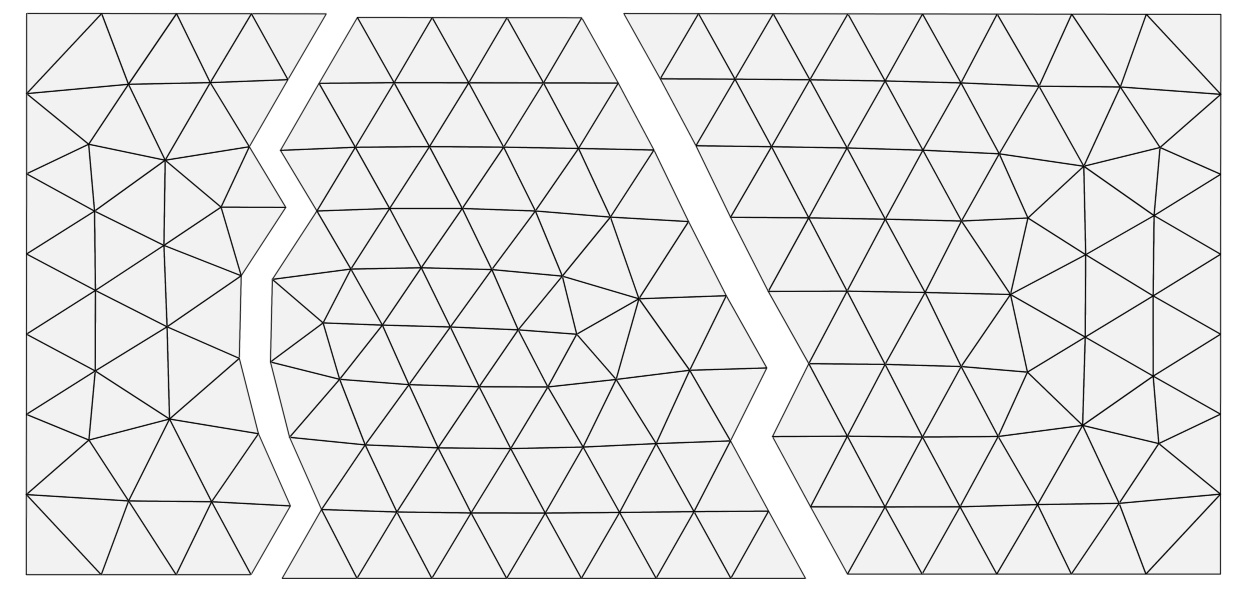}
 \includegraphics[width=0.49\linewidth]{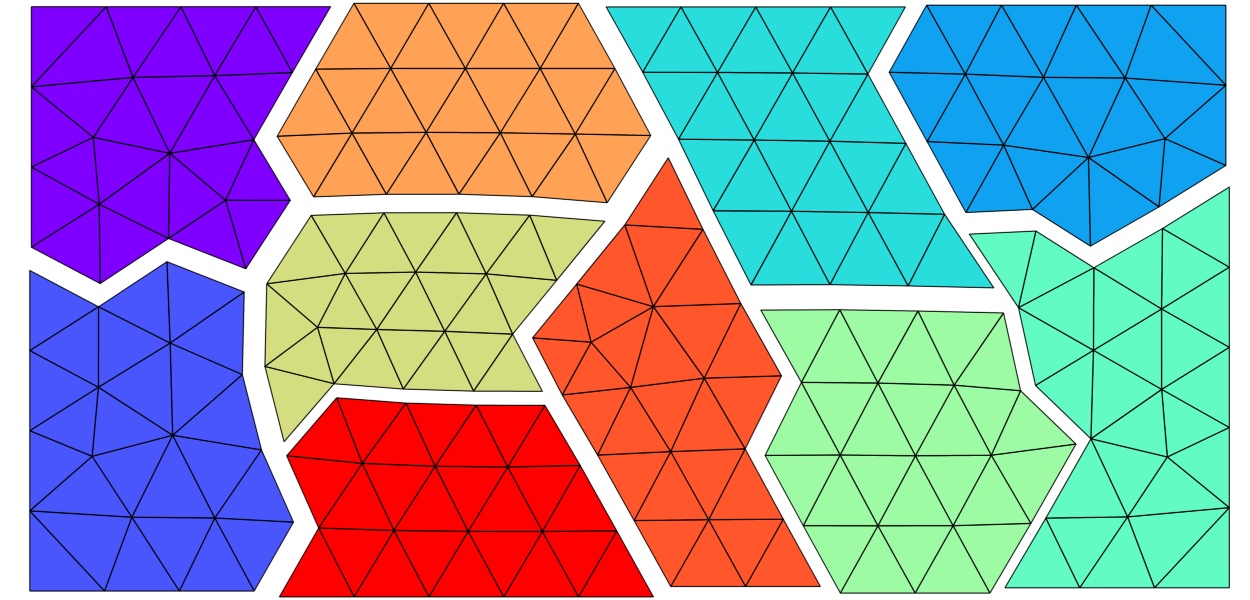} 
 \caption{Demonstration of partitioning a mesh into 10 submeshes. Each
   compute node has 4 processor cores, and 3 compute nodes are
   available. The first submesh is split into 2 small submeshes, and the
   second and third submeshes are partitioned into 4 small submeshes,
   respectively.\label{fig:hierarch_partition_2d_10}}
\end{figure}

Fig.~\ref{fig:hierarch_partition_2d_10} illustrates the process of
partitioning a mesh into $np$ processors when $np$ and $np_2$, the
number of processor cores on each compute node, are relatively prime.
The example assumes that the number of processor cores per compute
node is $np_2=4$, but $np=10$ submeshes are required. In the first
step of the algorithm, the mesh is partitioned into 3 subdomains of
unequal size, with the first subdomain being relatively smaller than
the other two. The first submesh is subsequently split into 2 smaller submeshes
while other two submeshes are each divided into 4 smaller submeshes. The
resulting 10 partitions are then of nearly equal size.

\section{Node assignment algorithm}
By default, libMesh \cite{libMeshPaper} assigns mesh nodes at partitioning
interfaces to the neighboring MPI process with the lower rank. This
simple heuristic can lead to a computational load imbalance in which
lower MPI ranks have more mesh nodes than higher MPI ranks, \emph{even when
  the element partition is perfectly balanced}.

An alternative, yet still simple and inexpensive, approach for
balancing the mesh nodes on each partition is to assign a node
randomly to an MPI rank if that rank has at least one of the node's
attached elements. Unfortunately, this simple approach may cause
preconditioner and solver applications to scale unpredictably, and is
not optimal for preserving the locality of an element and its
neighbor's data structures, which can be detrimental to the efficiency
of a finite element solver.

\begin{figure}
 \centering
 \includegraphics[width=0.3\linewidth]{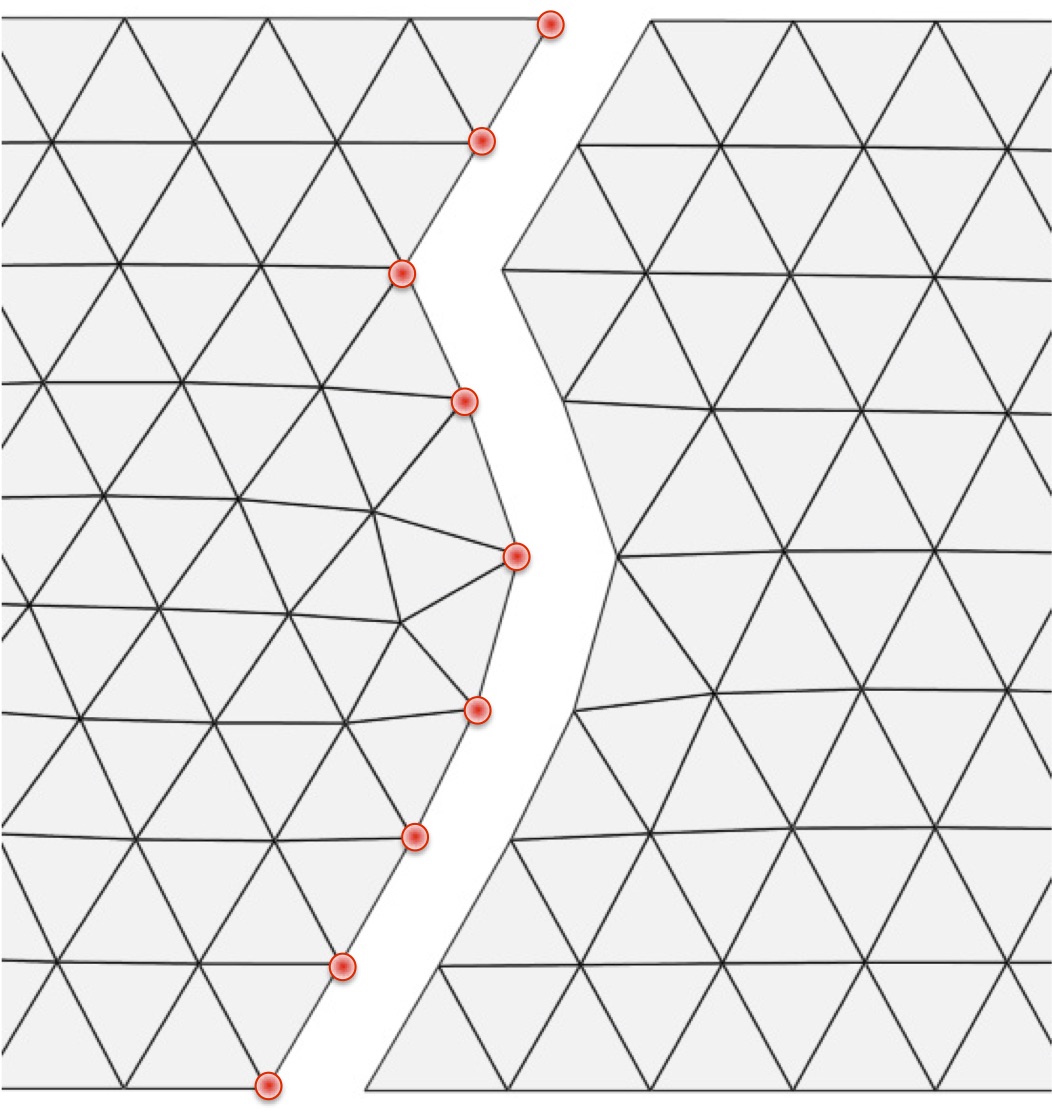}\quad
 \includegraphics[width=0.3\linewidth]{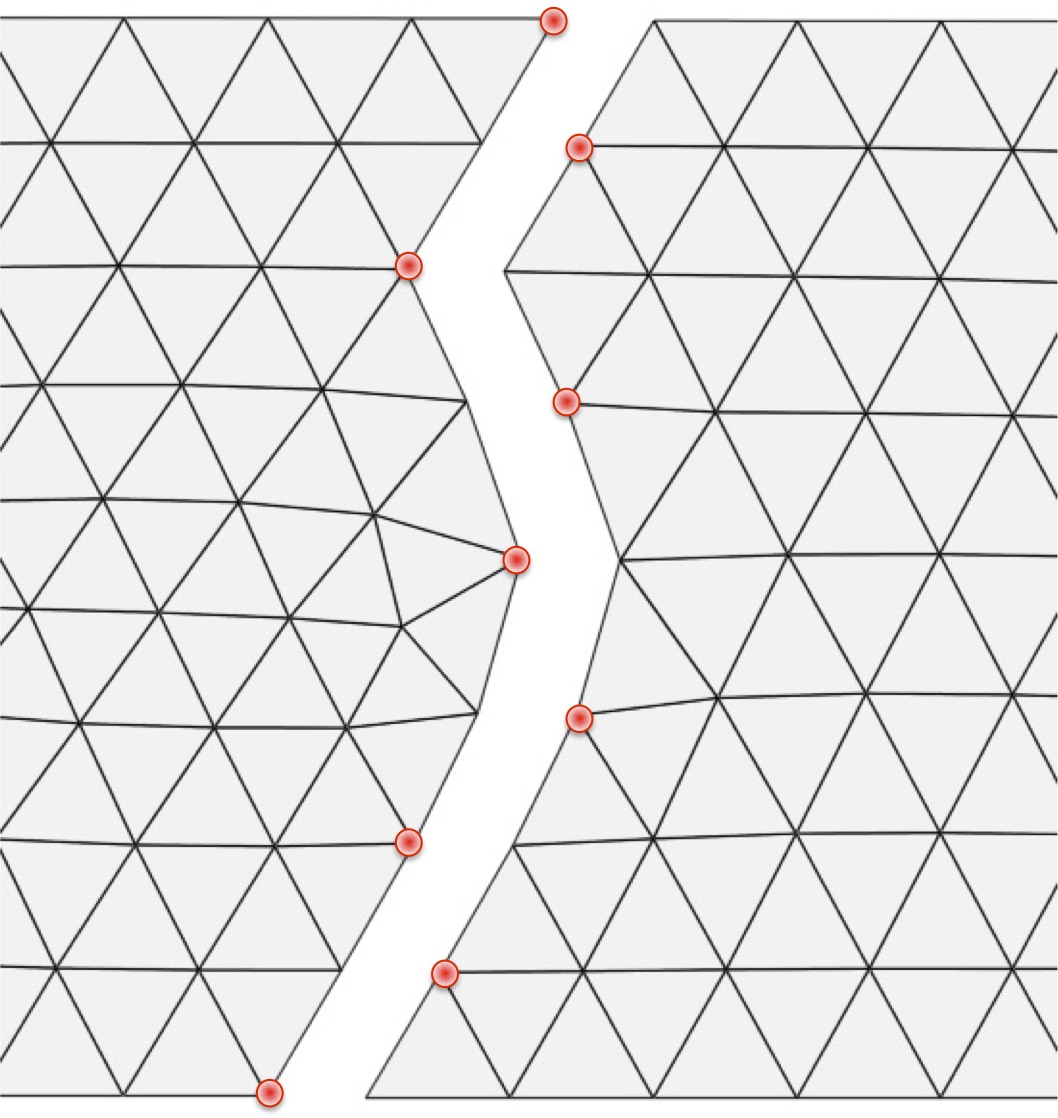}\quad
  \includegraphics[width=0.3\linewidth]{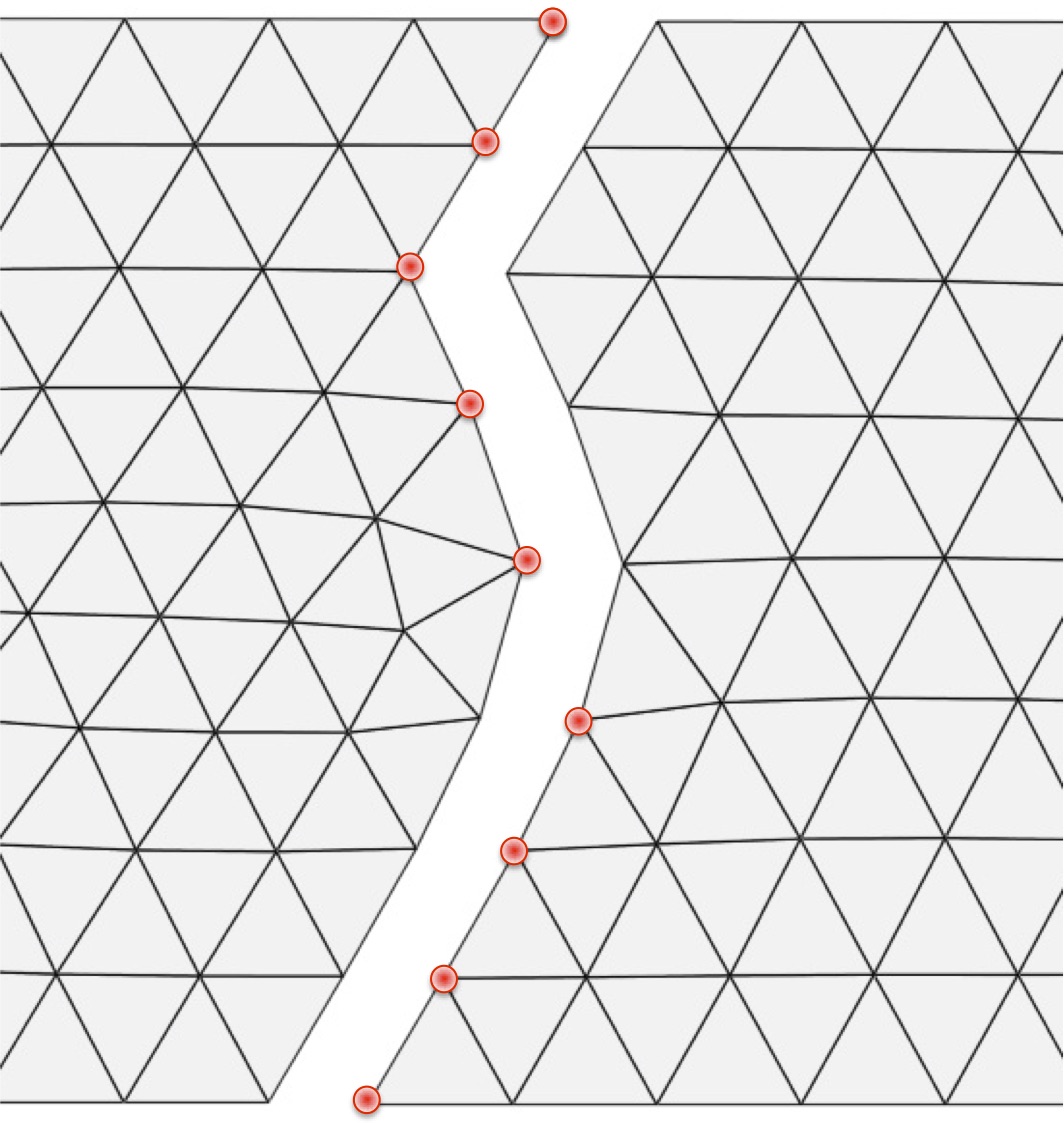}
  \caption{Different mesh node assignment schemes. Left: shared mesh nodes are
    assigned to the neighboring MPI process with the lower rank. Middle: mesh nodes with
    odd IDs are assigned to the lower MPI rank, even IDs are assigned to the
    higher MPI rank. Right: shared interface nodes are partitioned into two parts using a
    partitioning algorithm such as ParMetis or PTScotch. The approach on the right
    is the most efficient for the test case discussed in Sec.~\ref{sec:test_case}.\label{fig:nodeassignment}}
\end{figure}

We propose a novel algorithm to resolve this issue. The basic idea is
to apply a partitioner to each pair of MPI processes at the
lower-dimensional shared interface between processor boundaries, and
assign one resulting submesh to each neighboring MPI rank. Note that
each interface mesh is shared by only two MPI processes. The basic
idea is shown in Fig.~\ref{fig:nodeassignment} in detail. This new mesh
node assignment algorithm is implemented in \texttt{libMesh}.

\section{Test case\label{sec:test_case}}
In this section, we verify the effectiveness of the proposed
algorithms with a test case. The parallel preconditioning efficiency of
the proposed hierarchical partitioner is compared to
that obtained by applying ParMETIS and PTScotch directly. In
particular, the novel node assignment algorithm based on the interface
mesh partitioning will be shown to significantly improve
preconditioning performance.  Due to space constraints, we restrict
our testing to a single grain growth example here, but the proposed
 algorithm also works for other physical applications.

Grain growth is the increase in size of grains in a material due to a
reduction of the internal energy that is achieved by reducing the
total area of grain boundaries (GBs). GBs migrate to reduce the total
free energy of the system. Various sources of free energy drive the GB
migration process, including stored defect energy, deformation energy, and GB
energy. Various modeling approaches have been applied to model grain
boundary migration, and the phase field method has emerged as one of
the more popular. In the phase field model, each grain is represented by a
continuous order parameter $\eta_i$ that is equal to 1 within the grain and
equal to 0 in all other grains. The free energy for this problem is
\begin{align}
  f_{loc} = \mu \left( \sum_i^N \left(\frac{\eta_i^4}{4} - \frac{\eta_i^2}{2} \right)  + \gamma \sum_{i=1}^N \sum_{j>i}^N \eta_i^2 \eta_j^2 \right) + \frac{1}{4}
\end{align}
where $N$ is the number of order parameters, and $\mu,\gamma$ are
material coefficients. The $\eta_i$ evolve in space and time according
to the Allen-Cahn equation,
\begin{align}
  \label{eq:AC}
  \frac{\partial \eta_j}{\partial t} = - L_j \frac{\delta F}{\delta \eta_j},
\end{align}
where  $L_j$ is the order parameter mobility.  Here $F$ is defined as:
\begin{align}
    F = \int_V \big[ f_{loc}( \eta_1, \ldots, \eta_N) + f_{gr}( \eta_1,  \ldots, \eta_N)  \big] \, dV,
\end{align}
where the gradient energy density $f_{gr}$ is 
\begin{align}
  f_{gr} =  \sum^N_j \frac{\kappa_j}{2} |\nabla \eta_j|^2.
\end{align}
The model parameters $L_j$, $\mu$ and $\kappa_j$ are defined in terms
of the grain boundary (GB) surface energy $\sigma$, the diffuse GB
width $w_{GB}$ and the GB mobility $m_{GB}$. The values of the
parameters used in the present simulation are:
$L_j = 0.0354524$, $\mu=0.662848$,
$\gamma=1.5$ and $\kappa_j = 132.57$.

\begin{figure}
 \centering
 \includegraphics[width=0.49\linewidth]{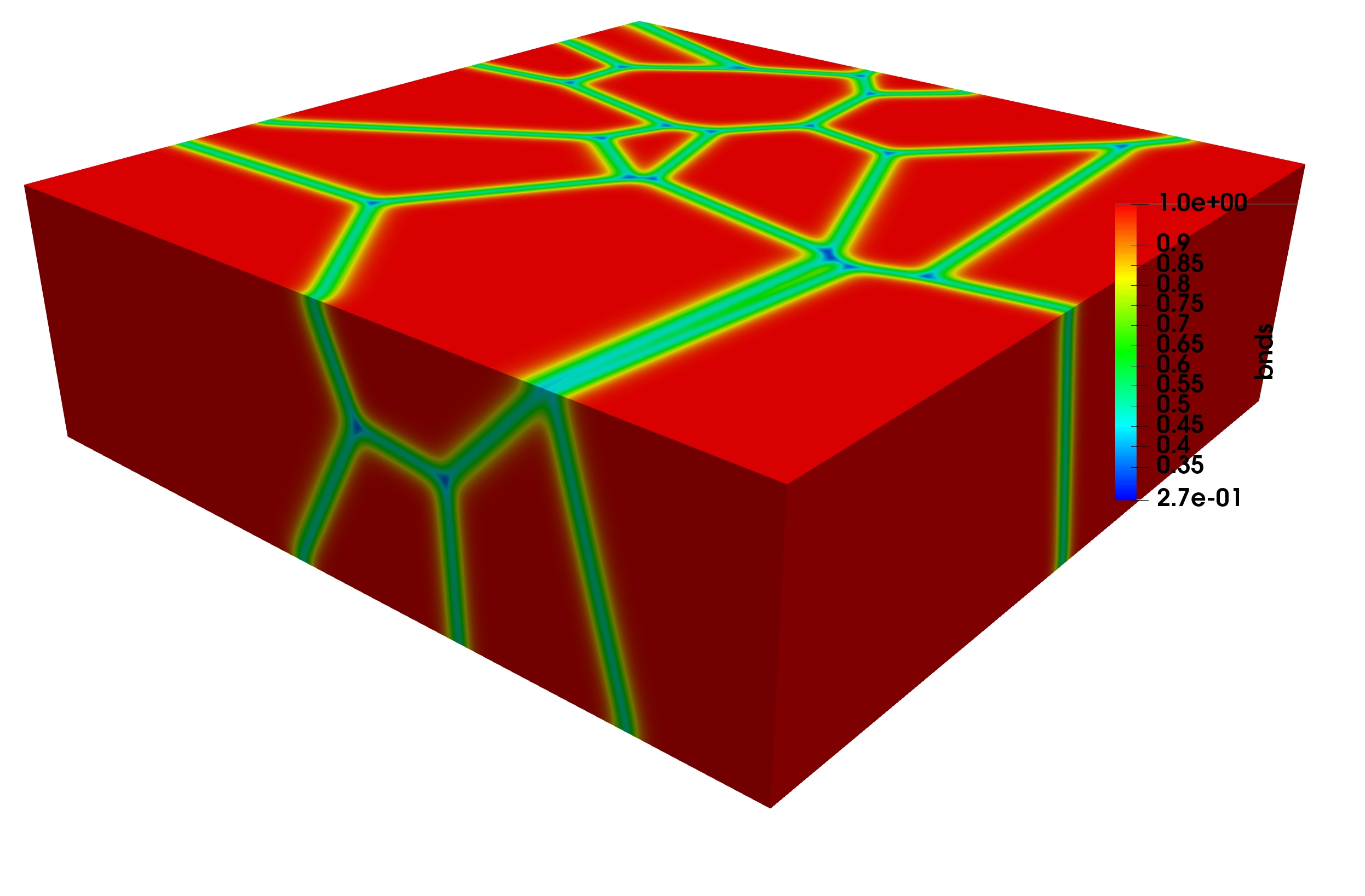}
 \includegraphics[width=0.49\linewidth]{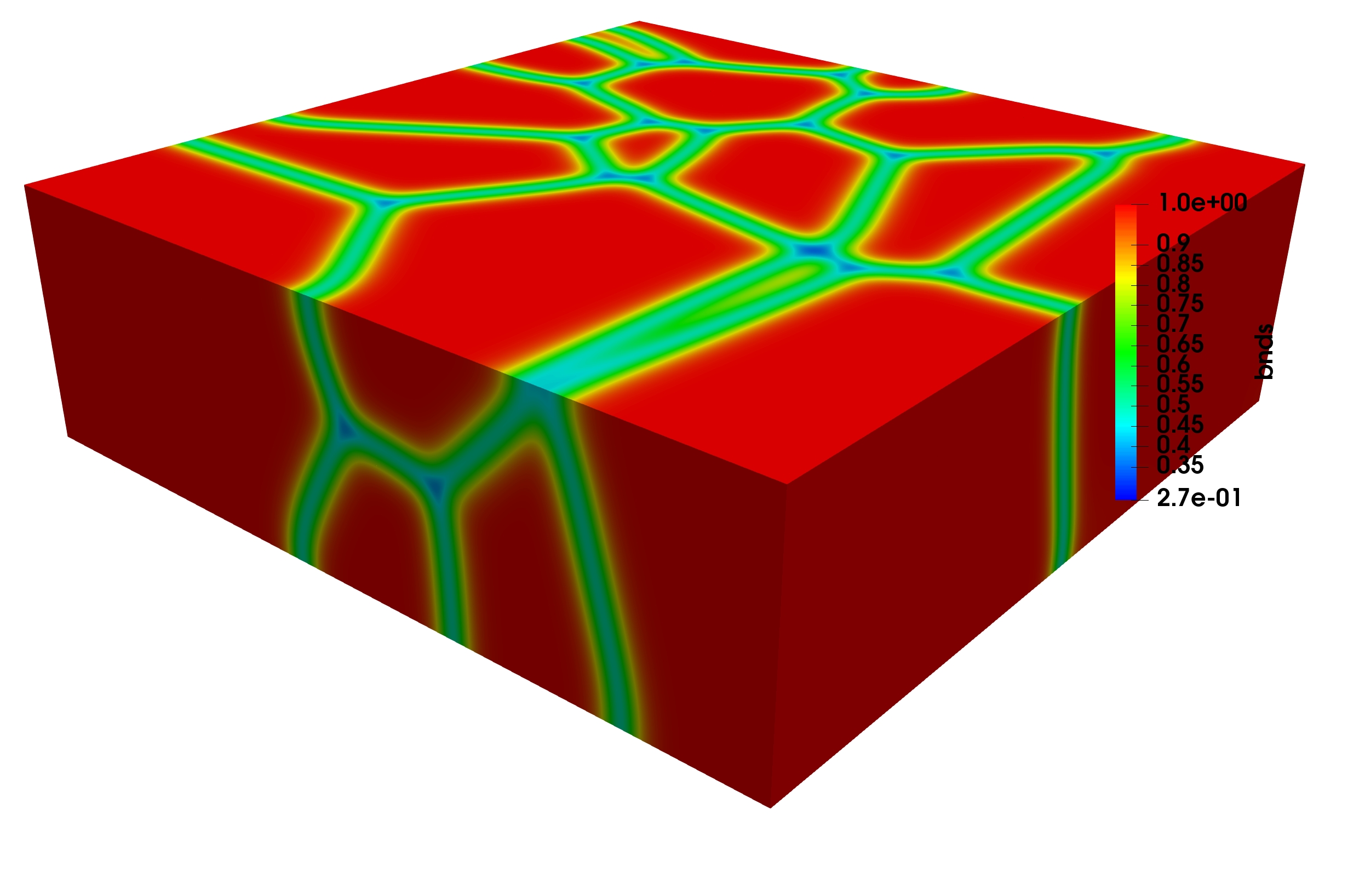} 
 \caption{Solution at $t=25$ns and $t=250$ns.\label{fig:grain_growth_time_stepper}}
\end{figure}

Eq.~\eqref{eq:AC} is discretized in 3D with $25$ grains and $9$ order
parameters using a first order Lagrange finite element method in MOOSE
\cite{Gaston_2015}. (The order parameters are reused for more than one
grain based on a coloring of the adjacency matrix representing the
grain connection of the microstructures.) The resulting nonlinear
system is solved using a Jacobian-free Newton--Krylov method
\cite{knoll2004jacobian}, employing GMRES \cite{saad1986gmres}
together with a restricted additive Schwarz preconditioner
\cite{cai1999restricted, kong2018fully}. The impact of the various
mesh partitioning methods on the preconditioning efficiency are
reported below. For reference, the solution at $t=25$ns and $250$ns is
shown in Fig.~\ref{fig:grain_growth_time_stepper}.

We initially solve this problem on a relatively ``coarse'' mesh
with 9,830,400 hexahedral elements, 9,994,977 nodes and 89,954,793
unknowns. The preconditioner performance for 10 time steps obtained
using various partitioning schemes is reported in
Table~\ref{tab:partitioning_coarse}. The columns of the table are
defined as follows: $np$ is the number of processor cores,
``EPart'' denotes the partitioner used for partitioning the mesh,
``PCSetup'' is the compute time in seconds spent on the preconditioner
setup, ``PCApply'' is the compute time in seconds spent on the
application of the preconditioner, ``PCSEFF'' is the parallel
efficiency of the preconditioner setup, and ``PCAEFF'' is the parallel
efficiency of the preconditioner application. ``NR'' is the ratio of
the maximum mesh node count to the minimum mesh node count;
``NR=1.0'' indicates the problem is perfectly balanced.

The test is conducted for processor counts between 8,192
and 32,768 on the Theta supercomputer at Argonne National
Laboratory. Theta is a massively parallel, many-core system with
second-generation Intel Xeon Phi processors. Each compute node has
a 64-core processor with 16 gigabytes (GB) of high-bandwidth
in-package memory (MCDRAM), 192 GB of DDR4 RAM, and a 128 GB SSD.

\begin{table}
\scriptsize
\centering
\caption{Partitioner performance on ``coarse'' mesh with 9,830,400
  elements. The times and parallel efficiencies for setting up
  and applying the preconditioner are measured individually, and the
  ratio of maximum to minimum node count is reported in the ``NR'' column.\label{tab:partitioning_coarse}}
\begin{tabular}{c c c c c c  c c c c}
\toprule
$np$  &EPart & PCSetup & PCApply & PCSEFF & PCAEFF  & NR \\
\midrule
8,192 & Hierarch  & 19.46 &8.1 &100\%& 100\%& 1.35 \\
8,192 & ParMETIS  & 22.76 & 10.38 &86\%&78\% & 1.57 \\
8,192 & PTScotch  & 20.33 &8.92 & *&* & 1.33 \\
\midrule
10,240 &  Hierarch  &16.21& 6.6& 96\%&  98\% & 1.45\\
10,240 &  ParMETIS  & 19.09& 8.17& 82\%&  79\% & 1.70\\
10,240 &  PTScotch  &*& *& *&  * & *\\
\midrule
16,384 & Hierarch  & 11.02& 5.14& 88\%  & 79\% & 1.54\\
16,384 & ParMETIS  & 13.2& 5.88& 74\% & 69\% & 1.94\\
16,384 &  PTScotch  &*& *& *&  * & *\\
\midrule
24,576 & Hierarch  & 8.71&  3.98 & 74\% & 68\% & 1.68\\
24,576 & ParMETIS  & 10.6&  4.8 &  61\% & 56\% & 2.04\\
24,576 &  PTScotch  &*& *& *&  * & *\\
\midrule
32,768 & Hierarch  & 6.8&  3.15 & 72\% & 64\% & 1.83\\
32,768 & ParMETIS  & 8.46&  4.35 & 58\% & 47\% & 1.97\\
32,768 &  PTScotch  &*& *& *&  * & *\\
\bottomrule
\end{tabular}
\end{table}

As shown in Table~\ref{tab:partitioning_coarse}, in the 8,192-core
case, PTScotch gives a result similar to the hierarchical partitioning
method, but it fails to partition the mesh for all other processor
counts, as indicated in the table with an asterisk. ParMETIS is able
to generate partitions for all cases, but the partition quality is not
as high as Hierarch's in the different metrics reported.

\begin{figure}[h]
\centering
  \includegraphics[width=0.49\linewidth]{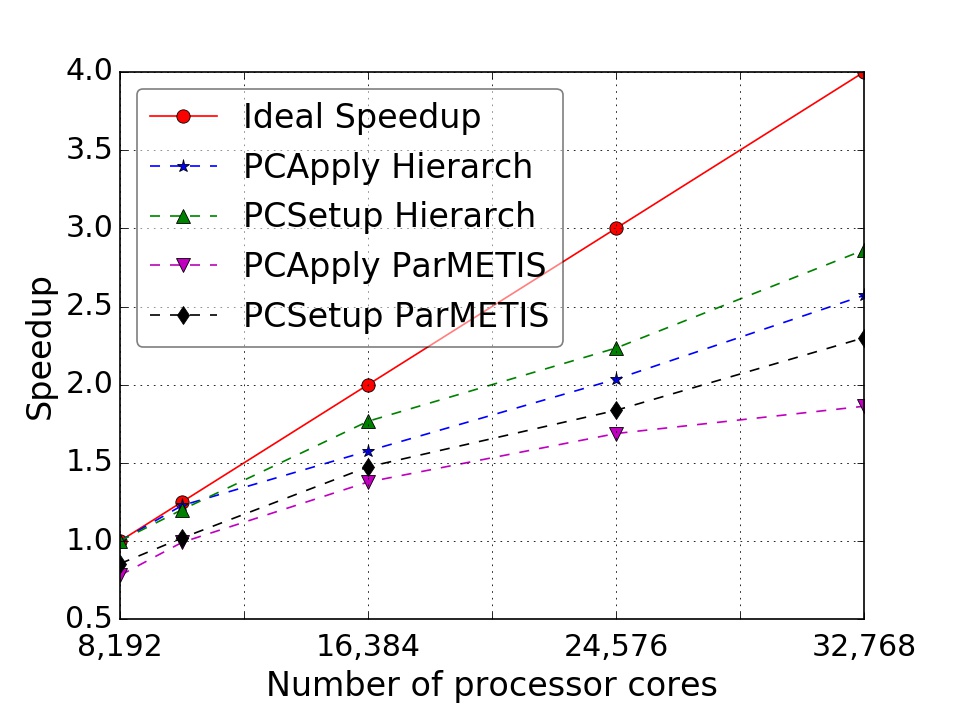} 
  \includegraphics[width=0.49\linewidth]{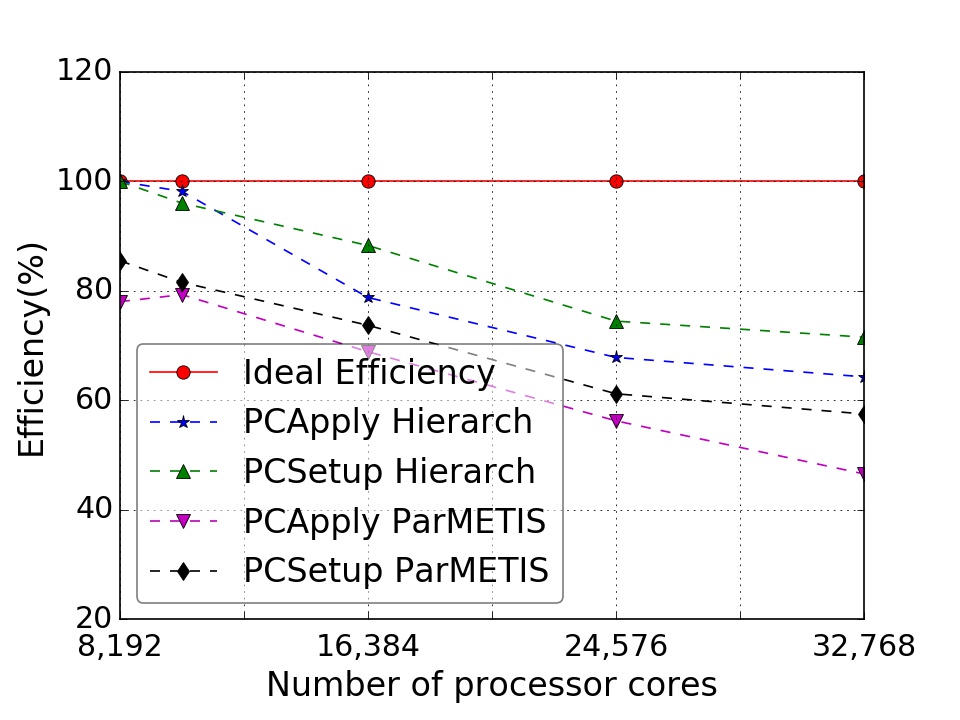} 
  \caption{Speedup and parallel efficiency for various partitioning approaches vs.\ core count
    on the ``coarse'' mesh problem.\label{fig:speedup_coarse_part}}
\end{figure}

The partition generated using Hierarch has better load balance, e.g.\
Hierarch's ``NR'' is 1.45 when using 10,240 processor cores while that
of ParMETIS is 1.7. The preconditioner application and setup times are
also shorter for Hierarch than for ParMETIS, despite the fact that the
same preconditioning algorithm was used in both cases. The parallel
efficiencies ``PCSEFF'' and ``PCAEFF'' for Hierarch are also 10-20
percentage points higher than the ParMETIS efficiencies, which indicates the proposed algorithm
scales better. The efficiencies and corresponding speedup are plotted
vs.\ core count in Fig.~\ref{fig:speedup_coarse_part}.

The results shown in Table~\ref{tab:partitioning_coarse} include the
new partitioner-based node assignment scheme discussed previously.  In
order to understand and isolate the importance of the node assignment
scheme to the overall preconditioner efficiency, we again solve the
``coarse'' problem with the hierarchical partitioner, but this time we
compare the ``default'' (minimum MPI rank) and partitioner-based node
assignment (``part'') schemes.

The numerical results are summarized in
Table~\ref{tab:nodeassignment_coarse}, and it is easily observed that,
in all cases, the new node assignment scheme significantly improves
the workload balance and reduces the total compute time. For example,
in the 8,192-core case, ``NR'' is 2.35 when using the default node
assignment scheme, while it is reduced to 1.35 when using the
partitioner-based node assignment algorithm.  Parallel efficiency is
generally improved by around $30\%$ when using the new node assignment
algorithm in place of the default one.

\begin{table}
\scriptsize
\centering
\caption{Effect of different node assignment algorithms. The
  ``coarse'' mesh problem is again solved using the hierarchical
  partitioner, but in this case we compare the ``default'' node assignment
  algorithm (minimum MPI rank) to the interface partitioner approach (``part'')
  for assigning node ownership.\label{tab:nodeassignment_coarse}}
\begin{tabular}{c c c c c c  c c c c}
\toprule
$np$  &NAS & PCSetup & PCApply & PCSEFF & PCAEFF  & NR \\
\midrule
8,192 & default  & 22.21 &13.5 &88\%&60\% & 2.35 \\
8,192 & part  & 19.46 &8.1 &100\%&100\% & 1.35 \\
\midrule
10,240 & default & 20.03 & 13.105 &78\%  & 49\% & 2.72\\
10,240 &  part  &16.21& 6.6& 96\%&  98\% & 1.45\\
\midrule
16,384 & default  &  13.01 & 8.6&75\% & 47\%& 3.04\\
16,384 & part  & 11.02& 5.14& 88\%  & 79\% & 1.54\\
\midrule
24,576 & default  & 9.4 & 6.9 &69\%& 39\% & 3.76\\
24,576 & part  & 8.71&  3.98 & 74\% &  68\%& 1.68\\
\midrule
32,768 & default  & 8.37 &  5.78 &58\%& 35\% & 4.1\\
32,768 & part  & 6.8&  3.15 & 71\% & 64\% & 1.83\\
\bottomrule
\end{tabular}
\end{table}

\begin{figure}
\centering
  \includegraphics[width=0.49\linewidth]{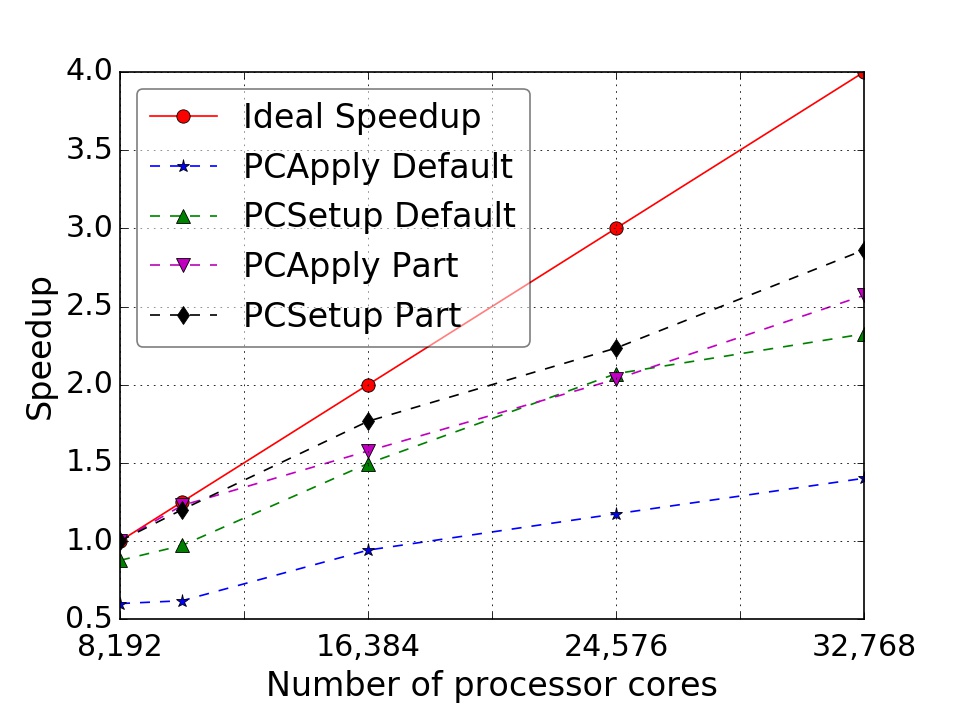} 
  \includegraphics[width=0.49\linewidth]{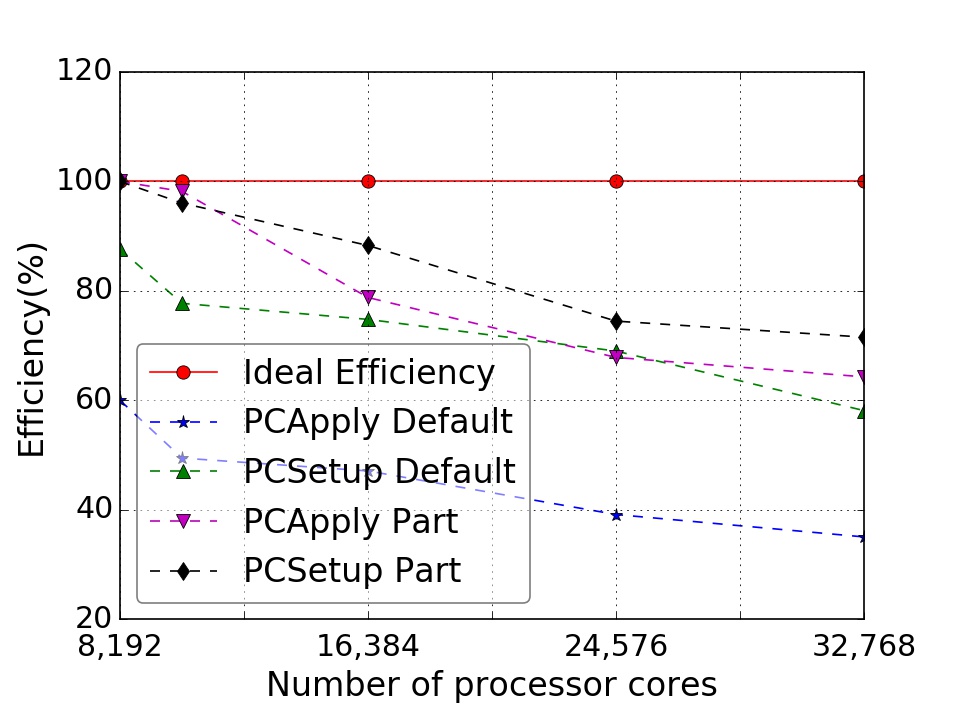} 
  \caption{Speedup and parallel efficiency for different node assignment algorithms vs.\ core count
    on the ``coarse'' mesh problem.\label{fig:speedup_coarse_node_assignment}}
\end{figure}

To further understand the scalability of the new partitioner, we also
ran the same problem on a ``fine'' mesh with 78,643,200 hexahedral elements
and 79,300,033 nodes. The resulting system has 713,700,297 unknowns,
and is again solved by the Jacobian-free Newton--Krylov method,
employing GMRES and a Schwarz preconditioner. Timings and parallel
efficiencies for the ``fine'' mesh are reported in
Table~\ref{tab:partitioning_fine}, where it is once again observed
that the hierarchical partitioner improves the preconditioner performance
significantly, as compared to ParMETIS. The PTScotch results are
omitted in this case since that partitioner was unable to generate a
partition of the fine mesh in any of the cases tested.

For the 8,192-core case, ``NR'' is only 1.14 (close to a perfect
balance ratio) when using Hierarch, while it is 2.4 when using
ParMETIS. The preconditioning process is $20\%$ faster for Hierarch
than it is for ParMETIS. The ParMETIS partitioner's performance can
also be irregular, as observed in the 24,576-core case where ``NR'' is
3.2 for ParMETIS and only 1.24 for Hierarch. This imbalance leads to
much slower preconditioner application and setup times (12s and 10s,
respectively) for the ParMETIS partitioner.

\begin{figure}
\centering
  \includegraphics[width=0.49\linewidth]{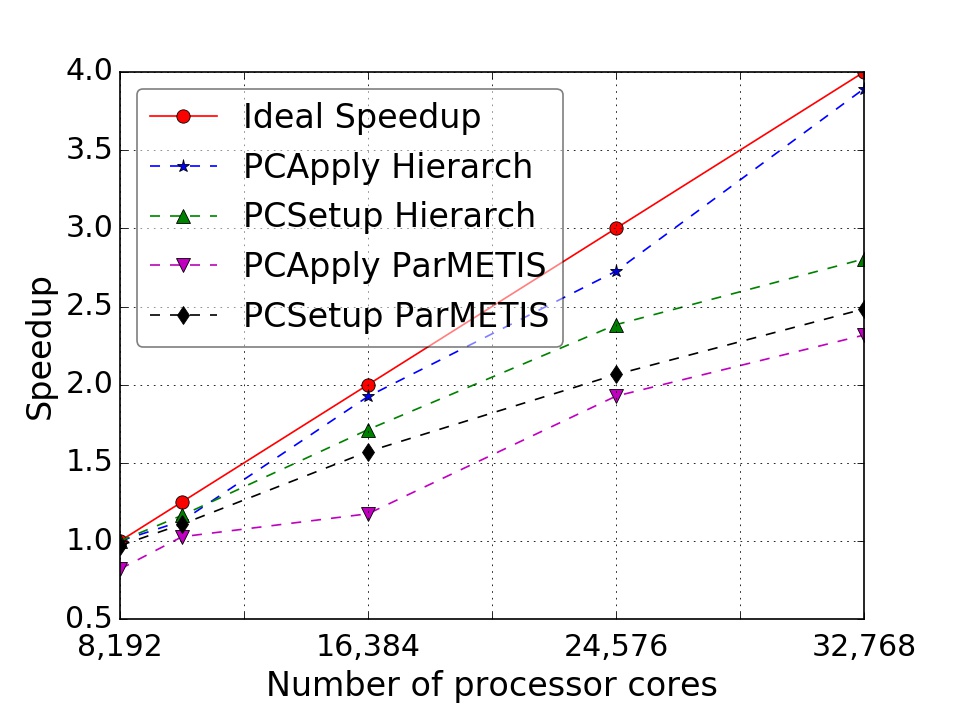} 
  \includegraphics[width=0.49\linewidth]{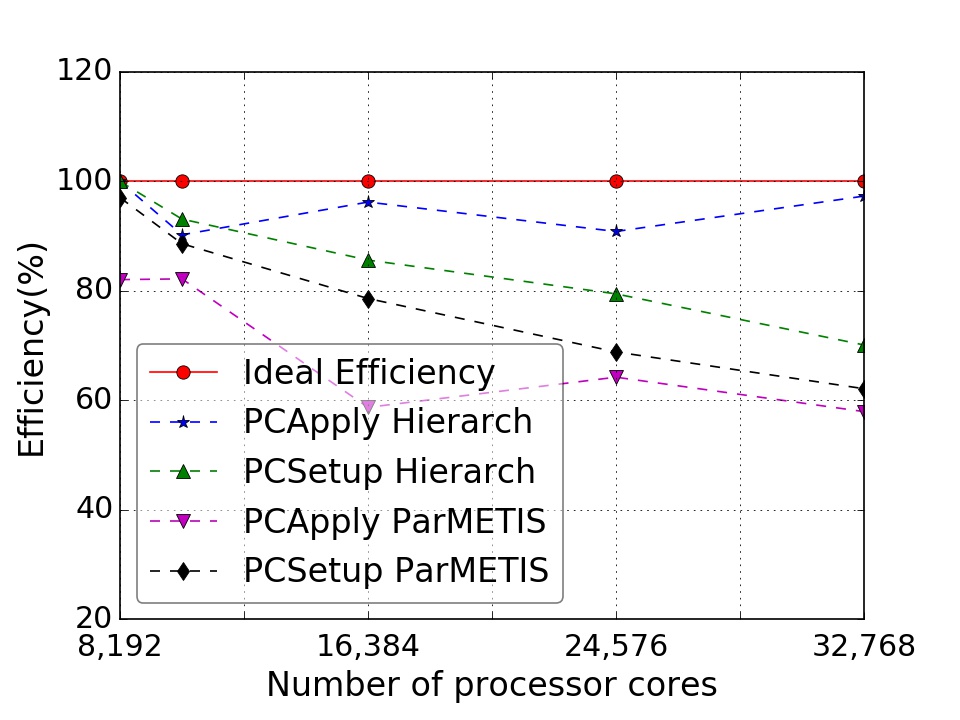} 
 \caption{Speedup and parallel efficiency for various partitioning approaches vs.\ core count
    on the ``fine'' mesh problem.\label{fig:speedup_fine_part}}
\end{figure}

The corresponding parallel efficiencies have a similar pattern, that
is, the parallel efficiency of the preconditioner application for
Hierarch is $30\%$ higher than that of ParMETIS, and the parallel
efficiency of the preconditioner setup for Hierarch is $10\%$ higher
than that of ParMETIS. The speedup and parallel efficiency for
the ``fine'' mesh case are also summarized in Fig.~\ref{fig:speedup_fine_part}.

\begin{table}
\scriptsize
\centering
\caption{Partitioner performance on ``fine'' mesh with 78,643,200 elements. PTScotch is
  omitted because it was not able to successfully generate any partitions of the fine
  mesh.\label{tab:partitioning_fine}}
\begin{tabular}{c c c c c c  c c c c}
\toprule
$np$  &EPart & PCSetup & PCApply & PCSEFF & PCAEFF  & NR \\
\midrule
8,192 & Hierarch  &  134.59 &80.65 &100\%&100\% & 1.14 \\
8,192 & ParMETIS  & 138.69 & 98.28 &97\%&82\% & 2.4 \\
\midrule
10,240 &  Hierarch  & 115.63& 71.494& 93\%&  90\% & 1.16\\
10,240 &  ParMETIS  & 121.50& 78.499& 88\%&  82\% & 1.87\\
\midrule
16,384 & Hierarch  & 78.59&  41.905& 86\%  & 96\% & 1.22\\
16,384 & ParMETIS  & 85.610& 68.675& 79\% & 59\% & 1.64\\
\midrule
24,576 & Hierarch  &  56.463&  29.589 & 79\% & 90\% & 1.24\\
24,576 & ParMETIS  & 65.17&  41.857 &  69\% & 64\% & 3.2\\
\midrule
32,768 & Hierarch  & 47.99&  20.717 & 70\% & 97\% & 1.34\\
32,768 & ParMETIS  & 54.165&  34.801 & 62\% & 58\%& 1.79\\
\bottomrule
\end{tabular}
\end{table}

\begin{figure}
\centering
  \includegraphics[width=0.49\linewidth]{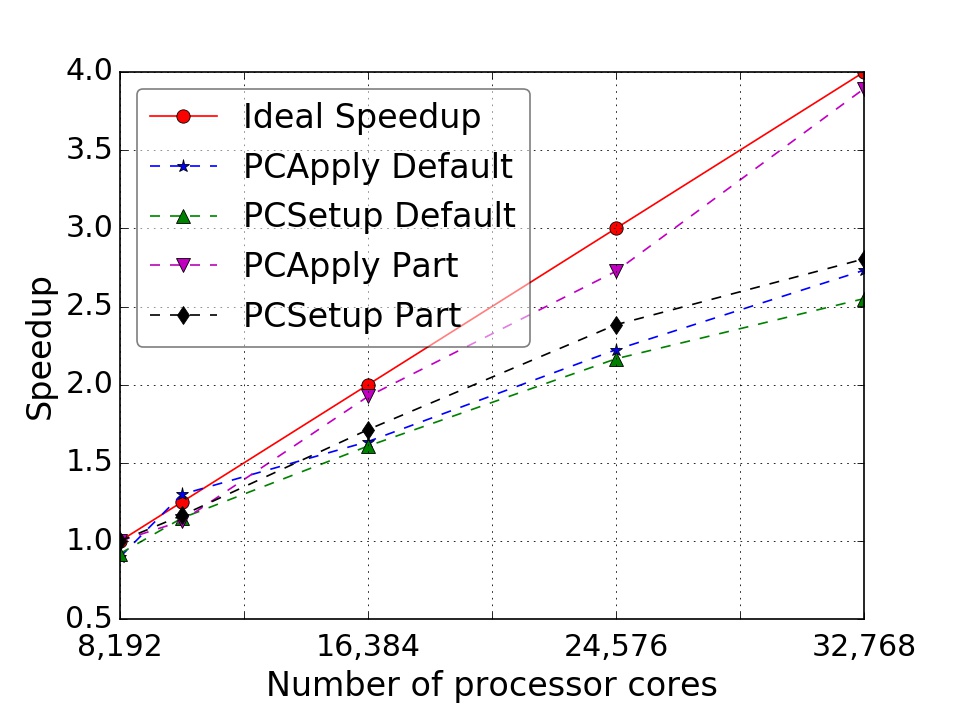} 
  \includegraphics[width=0.49\linewidth]{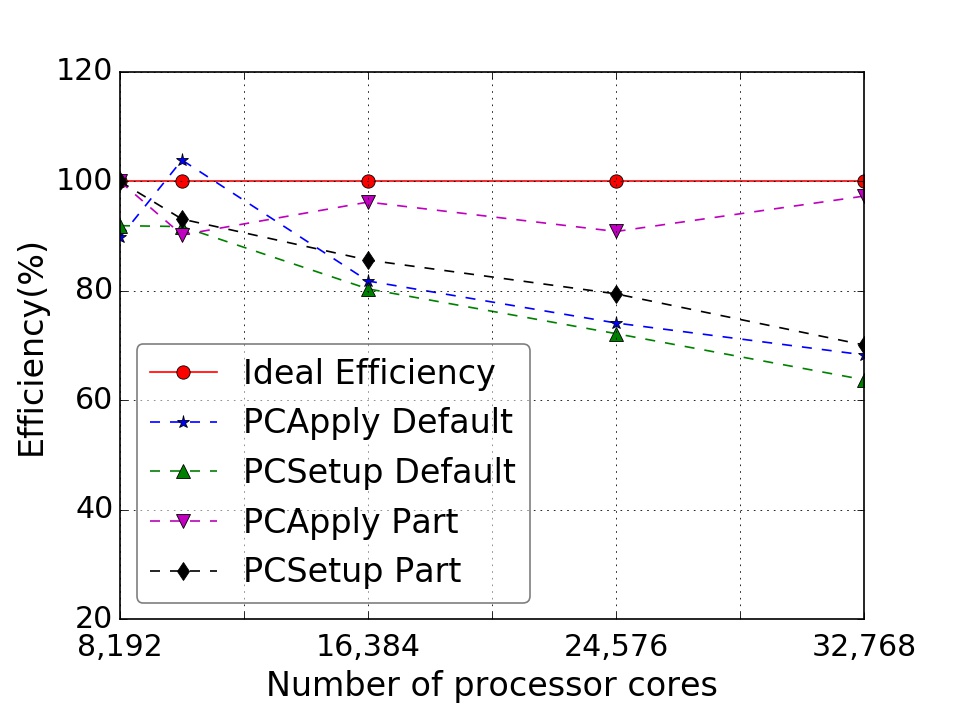} 
  \caption{Speedup and parallel efficiency for different node assignment algorithms vs.\ core count
    on the ``fine'' mesh problem.\label{fig:speedup_fine_node_assignment}}
\end{figure}

Similarly, we compare the preconditioning performance for the new node
assignment approach with using the ``default'' node assignment
heuristic.  We observe that, especially when the number of processor
cores is large, it is essential to maintain a workload balance in
order to achieve good parallel efficiency.  For example, when we use
32,768 processor cores, ``NR'' is 2.02 for the default node
assignment, while it is 1.34 for the partitioning-based node
assignment, and these different balance ratios lead to significantly
different overall performance levels.  The parallel efficiency of the
preconditioner application for the partitioner-based node assignment
is $20\%$ higher than that of the default node assignment algorithm,
while the parallel efficiency of the preconditioner setup is about
$10\%$ better.  The parallel efficiency and the speedup for the
different node assignment algorithms are also summarized in
Fig.~\ref{fig:speedup_fine_node_assignment}.

\begin{table}
\scriptsize
\centering
\caption{Effect of different node assignment strategies for the ``fine'' mesh. The slightly
  superlinear PCAEFF value in the 10,240 core case is sometimes observed at smaller
  core counts depending on the partitioner, but does not signify a general trend.\label{tab:nodeassignment_fine}}
\begin{tabular}{c c c c c c  c c c c}
\toprule
$np$  &NAS & PCSetup & PCApply & PCSEFF& PCAEFF  & NR \\
\midrule
8,192 & default  &  146.42 &  89.844 & 92\%&90\% & 1.51 \\
8,192 & part  &  134.59 &80.65 &100\%&100\% & 1.14 \\
\midrule
10,240 &  default  &  117.35 & 62.092& 92\%&  104\% & 1.56\\
10,240 &  part  & 115.63& 71.494& 93\%&  90\% & 1.16\\
\midrule
16,384 & default  & 83.736& 49.294& 80\% & 82\% & 1.72\\
16,384 & part  & 78.59&  41.905& 86\%  & 96\% & 1.22\\
\midrule
24,576 & default  & 62.149&  36.264 &  72\% & 74\% & 1.89\\
24,576 & part  &  56.463&  29.589 & 79\% &  91\%& 1.24\\
\midrule
32,768 & default  & 52.769&  29.520 & 64\% & 68\% & 2.02\\
32,768 & part  & 47.99&  20.717 & 70\% & 97\% & 1.34\\
\bottomrule
\end{tabular}
\end{table}

\section{Conclusions and future work}
A general-purpose hierarchical mesh partitioning method was introduced
and discussed for large-scale scientific computing. The partitioner
distributes graphs recursively onto both $np_1$ (the number of
compute nodes) and $np_2$ (usually $\leq$ the number of processor
cores per compute node) subdomains. Note that $np_2$ can be different
on each compute node, making the hierarchical partitioning algorithm
useful for general calculations on heterogeneous collections of nodes.

Mesh nodes on inter-processor interfaces are often assigned to the lower
MPI rank by default, and this simple choice can lead to a significant
load imbalance. The issue is addressed by introducing a new node
balancing algorithm in which a graph corresponding to the interface
mesh shared by two processor cores is partitioned into two
submeshes using a  partitioner, and one submesh is assigned to
the lower MPI rank while the other is sent to the higher MPI
rank. This scheme preserves the data locality and maintains a balance
in the computational workload. We numerically demonstrate that the
hierarchical partitioner approach combined with the new
node balancing technique reduces the application and setup time of the
preconditioner by about 50\% compared to direct application of
existing algorithms.

While this study focused on the correlation between partition quality
and preconditioner efficiency, future research will investigate the
performance and scalability of other parts of the simulation,
including the partitioning process itself.

\section*{Acknowledgments}
This research was supported through the INL Laboratory Directed Research \& Development (LDRD) Program under DOE Idaho Operation Office Contract DE-AC07-05ID14517 and used resources of the Argonne Leadership Computing Facility, which is a DOE Office of Science User Facility supported under Contract DE-AC02-06CH11357. This manuscript has been authored by Battelle Energy Alliance, LLC under Contract No. DE-AC07-05ID14517 with the U.S. Department of Energy. The United States Government retains and the publisher, by accepting the article for publication, acknowledges that the United States Government retains a nonexclusive, paid-up, irrevocable, and worldwide license to publish or reproduce the published form of this manuscript, or allow others to do so, for United States Government purposes.
\bibliographystyle{IEEEtran}
\bibliography{partitioning}

\end{document}